\documentclass[fleqn,usenatbib]{mnras}

\usepackage{newtxtext,newtxmath}

\usepackage{pifont}

\usepackage[T1]{fontenc}

\DeclareRobustCommand{\VAN}[3]{#2}
\let\VANthebibliography\thebibliography
\def\thebibliography{\DeclareRobustCommand{\VAN}[3]{##3}\VANthebibliography}

\usepackage{graphicx}	
\usepackage{amsmath}	
\usepackage{siunitx}
\usepackage{float}
\usepackage[normalem]{ulem}

\usepackage{booktabs, multirow} 
\usepackage{soul}
\usepackage[table]{xcolor} 
\usepackage{changepage} 

\usepackage{adjustbox} 
\usepackage{caption} 
\usepackage[para,online,flushleft]{threeparttable}
\usepackage{lipsum}

\newcommand{\rc}{$r$}
\newcommand{\rw}{$\Delta r$}
\newcommand{\rf}{$\Delta r$/$r$}

\newcommand{\smallspace}{\vspace{5pt}}

\title[Formation of wide exoKuiper belts]{The formation of wide \textit{exoKuiper} belts from migrating dust traps}

\author[]{
E. Miller,$^{1,2}$\thanks{E-mail: elle.ac.miller@gmail.com}
S. Marino,$^{3,4}$
S. M. Stammler,$^{5}$
P. Pinilla,$^{2,6}$
C. Lenz,$^{2}$
T. Birnstiel,$^{5,7}$,
and Th. Henning,$^{2}$
\\
$^{1}$Sydney Institute for Astronomy, University of Sydney, NSW 2006, Australia\\
$^{2}$Max-Planck-Institute f\"ur Astronomie, K\"onigstuhl 17, D-69117 Heidelberg, Germany\\
$^{3}$Jesus College, University of Cambridge, Jesus Lane, Cambridge CB5 8BL, UK\\
$^{4}$Institute of Astronomy, University of Cambridge, Madingley Road, Cambridge CB3 0HA, UK\\
$^{5}$University Observatory, Faculty of Physics, Ludwig-Maximilians-Universit\"at M\"unchen, Scheinerstraße 1, D-81679 Munich, Germany \\
$^{6}$Mullard Space Science Laboratory, University College London, Holmbury St Mary, Dorking, Surrey RH5 6NT, UK.\\
$^{7}$Exzellenzcluster ORIGINS, Boltzmannstr. 2, D-85748 Garching, Germany
}

\date{Accepted XXX. Received YYY; in original form ZZZ}

\pubyear{2021}

\begin{document}
\label{firstpage}
\pagerange{\pageref{firstpage}--\pageref{lastpage}}
\maketitle



\begin{abstract}
    The question of what determines the width of Kuiper belt analogues (\textit{exoKuiper} belts) is an open one. If solved, this understanding would provide valuable insights into the architecture, dynamics, and formation of exoplanetary systems. Recent observations by ALMA have revealed an apparent paradox in this field, the presence of radially narrow belts in protoplanetary discs that are likely the birthplaces of planetesimals, and \textit{exoKuiper} belts nearly four times as wide in mature systems. If the parent planetesimals of this type of debris disc indeed form in these narrow protoplanetary rings via streaming instability where dust is trapped, we propose that this width dichotomy could naturally arise if these dust traps form planetesimals whilst migrating radially, e.g. as caused by a migrating planet. Using the dust evolution software \textsc{DustPy}, we find that if the initial protoplanetary disc and trap conditions favour planetesimal formation, dust can still effectively accumulate and form planetesimals as the trap moves. This leads to a positive correlation between the inward radial speed and final planetesimal belt width, forming belts up to $\sim$100 au over 10 Myr of evolution. We show that although planetesimal formation is most efficient in low viscosity ($\alpha = 10^{-4}$) discs with steep dust traps to trigger the streaming instability, the large widths of most observed planetesimal belts constrain $\alpha$ to values $\geq4\times 10^{-4}$ at tens of au, otherwise the traps cannot migrate far enough. Additionally, the large spread in the widths and radii of \textit{exoKuiper} belts could be due to different trap migration speeds (or protoplanetary disc lifetimes) and different starting locations, respectively. Our work serves as a first step to link \textit{exoKuiper} belts and rings in protoplanetary discs.
    

\end{abstract}




\section{Introduction}

Over the last few years the Atacama Large Millimeter/submillimeter Array (ALMA) has revolutionised the study of circumstellar discs, both protoplanetary and debris discs. The unprecedented resolution and sensitivity of ALMA have revealed that the majority of large protoplanetary discs spanning tens of au are rich in substructure in the form of gaps and rings \citep[e.g.][]{Andrews2018, Long2018, Cieza2021}. The structure in those planet-forming discs is direct evidence of the presence of local pressure maxima stopping the radial drift of pebbles \citep[as once predicted by][]{whipple_1972} and trapping them in radially narrow rings, with typical fractional widths\footnote{The fractional width of a ring is defined as its width $\Delta r$ divided by its center position $r$.} $\lesssim0.2$ (Matr\'a et al. in prep). While the origin of these dust traps is not yet known, there are multiple scenarios which could explain their existence via the presence of already formed planets \citep[e.g.][]{Pinilla2012, Dipierro2015, Dong2017, Zhang2018, Perez2019}, or non-planet scenarios \citep{Pinilla2016, Flock2015, Takahashi2014, Loren-Aguilar2015, Saito2011, Dullemond2018a}.

On the other hand, ALMA has also shown that debris/planetesimal discs around mature systems that are massive analogues of the Kuiper belt, are not typically narrow as once thought \citep[e.g. as argued by][]{Strubbe2006}. The REsolved ALMA and SMA Observations of Nearby Stars (REASONS) survey \citep[][Matr\'a et al. in prep]{Sepulveda2019} has revealed that Kuiper belt analogues (or \textit{exoKuiper} belts) tend to be wide, with a median fractional width of 0.7. In fact, narrow debris rings such as Fomalhaut, $\epsilon$~Eri, HR~4796 and HD~202628 \citep{MacGregor2017, Booth2017, Kennedy2018, Faramaz2019} are rather rare and should not be considered as typical examples. This result comes as a surprise since the narrow dusty rings in protoplanetary discs are the ideal places to trigger planetesimal formation, e.g. regulated by streaming instability and gravitational collapse of pebble clouds \citep{Youdin2005, Johansen2007, Klahr_2020}, as shown recently by \citet{Stammler_2019} and \citet{Carrera2021}. Thus, it is an open question as to what makes debris discs wider if their planetesimals are born in these narrow rings. 

A possible explanation to this dichotomy is that the planetesimal formation in debris discs is regulated by streaming instability, but the formation is instead triggered at the final stages of protoplanetary disc evolution when surface densities drop due to photoevaporation and high dust-to-gas ratios are reached \citep[e.g.][]{Throop2005, Carrera2017, Ercolano2017}. This could naturally happen at a wide range of radii resulting in wide planetesimal belts. However, more recent simulations of discs undergoing photoevaporation do not find an increase in the dust-to-gas ratio, disfavouring this scenario \citep{Sellek2020}. Therefore, planetesimal formation at tens of au seems to require dust trapping in narrow rings. 

If the parent planetesimals of debris discs are indeed formed in those narrow rings in protoplanetary discs where dust is trapped, the apparent discrepancy in the width could be evidence for time-evolution of those rings. The underlying pressure maxima could appear and disappear at different radii if caused by transient phenomena \citep[e.g. vortices or zonal flows,][]{Johansen_2009, Uribe_2011, Flock2015}, which could lead to planetesimal formation from 1 to 50~au as shown by \cite{Lenz2019}. On the other hand, the traps could  migrate in radius \citep[e.g. if caused by a planet,][]{Li2009, Meru2019, Nazari2019, Shibaike2020}, and in this work we focus on this scenario to investigate the conditions in which moving dust traps could lead to the formation of wide planetesimal belts. While there are similarities between our aim and recent simulations by \cite{Shibaike2020}, they mainly focused on the disc regions between 1 to 30 au, i.e. interior to the typical distance at which \textit{exoKuiper} belts are found. Here we focus on planetesimal formation at distances between 10 and 150 au that overlay with the \textit{exoKuiper} belt population. Moreover, we use state-of-the-art simulations that compute dust coagulation and evolution in a viscously evolving gaseous disc, whereas \cite{Shibaike2020} used a more simple 1D Lagrangian particle model with pebbles with a single Stokes number. 

One additional scenario was proposed recently by \cite{Jiang2021}. They showed that a clumpy dust ring can form in a smooth gas disc and actively form planetesimals via streaming instability without dust trapping. In that scenario, pebbles drifting inwards encounter a small initial over-density in the pebble distribution that grows. This triggers the formation of dusty clumps in the midplane via streaming instability, and then these clumps undergo gravitational collapse forming planetesimals. This clumpy ring could move outwards due to the diffusion of clumps, which could lead to planetesimal formation at a wide range of radii, possibly explaining the width of \textit{exoKuiper} belts as well. One piece of evidence against this scenario is that multiple discs show kinematic deviations from Keplerian rotation at the location of dusty rings and gaps \citep[e.g.][]{Teague2018kinematics, Teague2018AS209, Teague2019}. Such deviations indicate perturbations in the pressure gradient that are consistent with local pressure maxima and dust trapping. In addition, although \cite{Jiang2021} show that the clumpy rings can migrate at speeds of up to $\sim10$~au~Myr$^{-1}$, and thus should lead to the formation of wide planetesimal belts over a disc lifetime, their migration speeds are still very uncertain due to the unknown drift velocity of clumps. Nevertheless, the clumpy ring scenario could explain some of the observed rings and perhaps wide \textit{exoKuiper} belts and deserves further investigation.

This paper is structured as follows. In \S\ref{sec:methods} the methods and model used to simulate the disc evolution and planetesimal formation is outlined. The primary simulation results are presented in \S\ref{sec:results} which determines the conditions favourable for planetesimal formation, and shows the outcomes of implementing these for a migrating dust trap. In \S\ref{sec:discussion} we examine the relationship between migration velocity and resulting planetesimal belt width, and explore the effect of several simulation parameters. The main findings of the paper are then summarised in \S\ref{sec:conclusion}. More detailed results including the mass evolution and dust distribution in the simulations can be found in the Appendix.

\section{Methods}
\label{sec:methods}
To simulate the multiple processes governing the dust and gas evolution we use the software package \textsc{DustPy}\footnote{stammler.github.io/dustpy}, created by Stammler \& Birnstiel (in prep). \textsc{DustPy} uses numerical methods to evolve a protoplanetary system based upon the model by \cite{Birnstiel_2010}, considering viscous evolution, dust coagulation and fragmentation, dust advection and diffusion. In our simulations, each system is evolved for 10 Myr. This is an approximate upper limit of protoplanetary disc lifespan  \citep[e.g.][]{Haisch_Jr__2001, fedele_2010, ribas_2015}, but it could also be higher \citep{Pfalzner_2014, Michel2021}. Below we describe the disc model in more detail.

\subsection{Disc model}
We consider a disc with a Solar-type star at its center. The input parameters to the disc model detailed in Table~\ref{tab:ic} were fixed for all simulations (except in specified exploratory situations). The only disc parameter varied is $\alpha$, which describes the efficiency of angular momentum transport due to turbulence. Two $\alpha$ cases were explored in each dust trap scenario. 

The radial grid of the disc is logarithmically spaced from 10 to 250 au in 240 bins. For $\alpha = 10^{-3}$ discs the dust mass distribution grid is spaced from $10^{-12}$ to $10$ grams with 7 mass bins per decade, totalling 92 bins. For $\alpha = 10^{-4}$ discs the grid is spaced from $10^{-12}$ to $10^{8}$ grams with 7 mass bins per decade, totalling 141 bins. The $\alpha = 10^{-4}$ discs are set with a larger maximum grain mass due to the higher fragmentation limit. Note that since dust will only reach this limit in the inner regions, this does not have an impact on the main results.


\subsection{Temperature profile}
The mid-plane gas temperature profile is assumed to follow a passively irradiated disc, given by
\begin{equation}
    T(r) = \Bigg(\frac{T_\star^4 R_\star^2 \phi }{r^2} \Bigg)^\frac{1}{4}
     = T_\star (R_\star \sqrt{\phi})^\frac{1}{2} r^{-\frac{1}{2}} \approx 260 \Bigg(\frac{1 \rm au}{r} \Bigg)^\frac{1}{2} {\rm K}.  \label{eq:temp}
\end{equation}

It is assumed the gas and dust have the same temperature, and that the disc is vertically isothermal. The stellar parameters are fixed throughout the simulation.

\begin{table}
 \begin{adjustbox}{max width=1.0\columnwidth}

   \begin{threeparttable}
	\caption{Initial conditions of the disc model.}
	\label{tab:ic}
	\centering
	\begin{tabular}{llll}
		\hline
		Parameter & Description & See Eq. & Value(s) \\
		\hline
		$\alpha$ & turbulent viscosity efficiency & \ref{eq:kin}, \ref{eq:alpha}, \ref{eq:vab}, \ref{eq:fab} & \{$10^{-3}, 10^{-4}$\}\\
		$A$ & gap amplitude & \ref{eq:bump} & \{3, 10\} \\
		$r_{\rm g}$ & gap position & \ref{eq:bump} & \{30, 60, 90 au\} \\
		$f$ & gap velocity nominal fraction & \ref{eq:fab} & \{0.1, 0.3, 1, 3\}\\
		$r_{\rm min}$ & disc minimum radius & & 10 au \\
		$r_{\rm max}$ & disc maximum radius & & 250 au \\
	    $\Sigma_{\rm d}/\Sigma_{\rm g}$ & initial dust-to-gas ratio & & 0.01 \\
	    $M_{\rm disc}$ & initial disc mass & \ref{eq:gsd} & 0.1 M$_\odot$ \\
	    $M_\star$ & stellar mass & \ref{eq:vab}, \ref{eq:B} & 1 M$_\odot$\ \\
	    $R_\star$ & stellar radius & \ref{eq:temp}, \ref{eq:vab} & 2 R$_\odot$ \\
		$T_\star$ & stellar temperature & \ref{eq:temp}, \ref{eq:vab} & 5772 K \\
		$\phi$ & irradiation angle & \ref{eq:temp}, \ref{eq:vab} & 0.05 rad \\

		$\rho_{\rm s}$ & solid density of dust grains & \ref{eq:stokes}, \ref{eq:vf} & 1.6 g~cm$^{-3}$\\
		
		$\mu$ & mean molecular weight & \ref{eq:cs} & 2.3 \\
		
		$a_{\rm 0}$ & minimum initial grain size & & \SI{1}{\micro\metre}\\
		$v_{\rm f}^{a}$ & fragmentation velocity & \ref{eq:vf} & 10 m~s$^{-1}$ \\
		$\delta_{\rm r,z,t}^{a}$ & radial diffusion, settling  & \ref{eq:D}, \ref{eq:dr}, \ref{eq:dt}, \ref{eq:vf} &  $\alpha$\\
                        & and turbulence parameters & \\
                        
        $n^{b}$ & smoothness parameter & \ref{eq:pf_prob} & 0.03\\
        $\zeta^{c}$ & planetesimal formation efficiency & \ref{eq:pfrate} & 0.1\\

		\hline
	\end{tabular}
	\begin{tablenotes}
	$^{a}$: In \S\ref{sec:vfrag} we explore lower values of $v_{\rm f}$ and values of $\delta$ different from $\alpha$.\\
	$^{b}$: In \S\ref{sec:sharp} we explore other smoothness values. \\
	$^{c}$: In \S\ref{sec:zeta} we explore various formation efficiencies. \\
	\end{tablenotes}
\end{threeparttable}  
 \end{adjustbox}

\end{table}

\subsection{Evolution of gas distribution}

We use an initial gas surface density profile following a self-similar solution of a viscously evolving disc \citep{LyndenBell1974TheEO}
\begin{equation}
    \Sigma_{\rm g} (r) = \Sigma_0 \bigg(\frac{r}{r_{\rm c}}\bigg)^{-\gamma} \exp{\Bigg[ -\bigg(\frac{r}{r_{\rm c}}\bigg)^{2-\gamma}\Bigg]},
    \label{eq:gsd}
\end{equation} 
where $r_{\rm c}$ = 60 au, $\gamma$ = 1 and $\Sigma_0$ is calculated to normalise the total gas mass according to $M_{\rm disc}$. 

The gas surface density $\Sigma_{\rm g}$ viscously evolves following the equations for diffusion, angular momentum conservation of the gas and assuming rotation at the Keplerian frequency \citep[e.g.][]{Pringle_1981}. Namely, its evolution is set by
\begin{equation}
    \frac{ \partial \Sigma_{\rm g} (r,t)}{\partial t} = \frac{3}{r}\frac{\partial}{\partial r}\left[ r^{1/2}\frac{\partial}{\partial r} (\nu \Sigma_{\rm g} r^{1/2})\right] ,
\end{equation}
where $\nu$ is the kinematic viscosity. Furthermore, we use the \cite{Shakura_1973} parametrisation, i.e. 
\begin{equation}
    \nu = \alpha c_{\rm s} H,
    \label{eq:kin}
\end{equation}
where $c_{\rm s}$ and $H$ are the isothermal sound speed and gas pressure scale height, respectively. The parameter $\alpha$ is assumed constant across the disc, except for a narrow range of radii where we perturb it to create a gap in the gas (see \S\ref{sec:peturb}). The value of $\alpha$ is uncertain, but observations of protoplanetary discs over the last five years have constrained it to values $\lesssim 10^{-3}$ at tens of au by studying the non-thermal broadening of emission lines \citep{Flaherty2015, Teague2016, Flaherty2017, Teague2018turbulence, Flaherty2020}, with some exceptions \citep[e.g.][]{Flaherty2020}.

\subsection{Evolution of dust distribution}
\label{sec:dust}
The initial dust surface density follows the same profile as the gas distribution, but considering the initial dust-to-gas mass ratio. Each dust particle in mass bin $i$ obeys its own advection-diffusion equation
\begin{equation}
    \frac{\partial}{\partial t}\Sigma_{\rm d}^i + \frac{1}{r} \frac{\partial}{\partial r}(r\Sigma_{\rm d}^i v^i_{\rm d,r}) = \frac{1}{r}\frac{\partial}{\partial r}\Big[rD^i \Sigma_{\rm g} \frac{\partial}{\partial r}\Big(\frac{\Sigma_{\rm d}^i}{\Sigma_{\rm g}}\Big)\Big].
\end{equation}
The radial dust velocity for each particle $i$ under gas pressure $P$ is
\begin{equation}
    v^i_{\rm d,r} = \frac{1}{1+\text{St}^{i^2}}v_{\rm g,r} + \frac{1}{\text{St}^i + 1/\text{St}^i} \frac{c_{\rm s}^2}{\Omega_{\rm K} r} \frac{{\rm d} \ln{P}}{ {\rm d} \ln{r}},
\end{equation}
where St is the Stokes number, $P$ the gas pressure and $\Omega_{\rm K}$ the Keplerian angular velocity. The dust diffusitivity $D$ is given by \cite{Youdin2007ParticleSI}
\begin{equation}
    D^i = \frac{\delta_{\rm r} c_s^2}{\Omega_{\rm K}(1 + \text{St}^{i^2})}.
    \label{eq:D}
\end{equation}
The Stokes number in the midplane is defined by \citep{Brauer_2008}
\begin{equation}
    \text{St}^i = \frac{\pi}{2} \frac{a^i\rho_{\rm s}}{\Sigma_g}.
    \label{eq:stokes}
\end{equation}
In principle, there are three values of $\delta$. $\delta_r$ influences the radial diffusion of the dust, as shown in Equation~\ref{eq:D}. $\delta_z$ influences the vertical setting, and used to calculate the dust scale heights $H_i$ as \footnote{stammler.github.io/dustpy/4\_standard\_model.html\#Simulation.dust.H}
\begin{equation}
    H_i = H \sqrt{\frac{\delta_{\rm r}}{\delta_{\rm r}+{\rm St}_i}}.
    \label{eq:dr}
\end{equation}
$\delta_{\rm t}$ impacts the turbulent collision velocities, which is approximated by \citep{OrmelCuzzi}
\begin{equation}
    v_{\rm f}^2 \sim 3\delta_{\rm t } c_{\rm s}^2 \text{ }{\rm St }.
    \label{eq:dt}
\end{equation}
See \cite{Pinilla_2021} for a more detailed description and discussion on these parameters. For simplicity, we assume the three have the same value across the disc and are equal to $\alpha$. Note that as explored recently by \cite{Pinilla_2021}, in principle the values of $\delta_{\rm  r,z,t}$ do not need to be isotropic and can be different from $\alpha$, which can have important consequences for the dust distribution. This possibility is discussed in \S\ref{sec:vfrag}.

\subsection{Dust growth and fragmentation}
\label{sec:fragmentation}
Dust grain growth and fragmentation were determined by solving the Smoluchowski equation \citep{smol}
\begin{equation}
    \frac{\partial}{\partial t} f(m) = \int \int f(m') f(m'') M(m, m', m'') dm'' dm'
\end{equation}
where the coagulation Kernel $M(m,m',m'')$ hides the exact collisional physics, see \cite{Birnstiel_2010} for details. Dust grains grow through
hit-and-stick collisions until their relative collision velocities exceed the fragmentation velocity $v_{\rm f}$. In our model we assume a fragmentation velocity of 10~m s$^{-1}$. This value is motivated by previous laboratory experiments and numerical simulations that investigated the sticking properties of water-ice particles \cite{Blum2000, Wada2009, Wada2011, Gundlach2015, Musiolik2016}, which would approximate well the properties of grains in our region of interest at tens of au from the star. Nonetheless, the fragmentation velocity remains highly uncertain and recent experiments have suggested that amorphous water ice could be significanltly less sticky \citep{gundlach_2018, Musiolik_2019, Steinpilz_2019}, and therefore  in \S\ref{sec:vfrag} we discuss the effect of lower values of $v_{\rm f}$. 

\subsection{Planetesimal formation}
\label{sec:prob}
\begin{figure}
    \centering
    \includegraphics[width=\columnwidth]{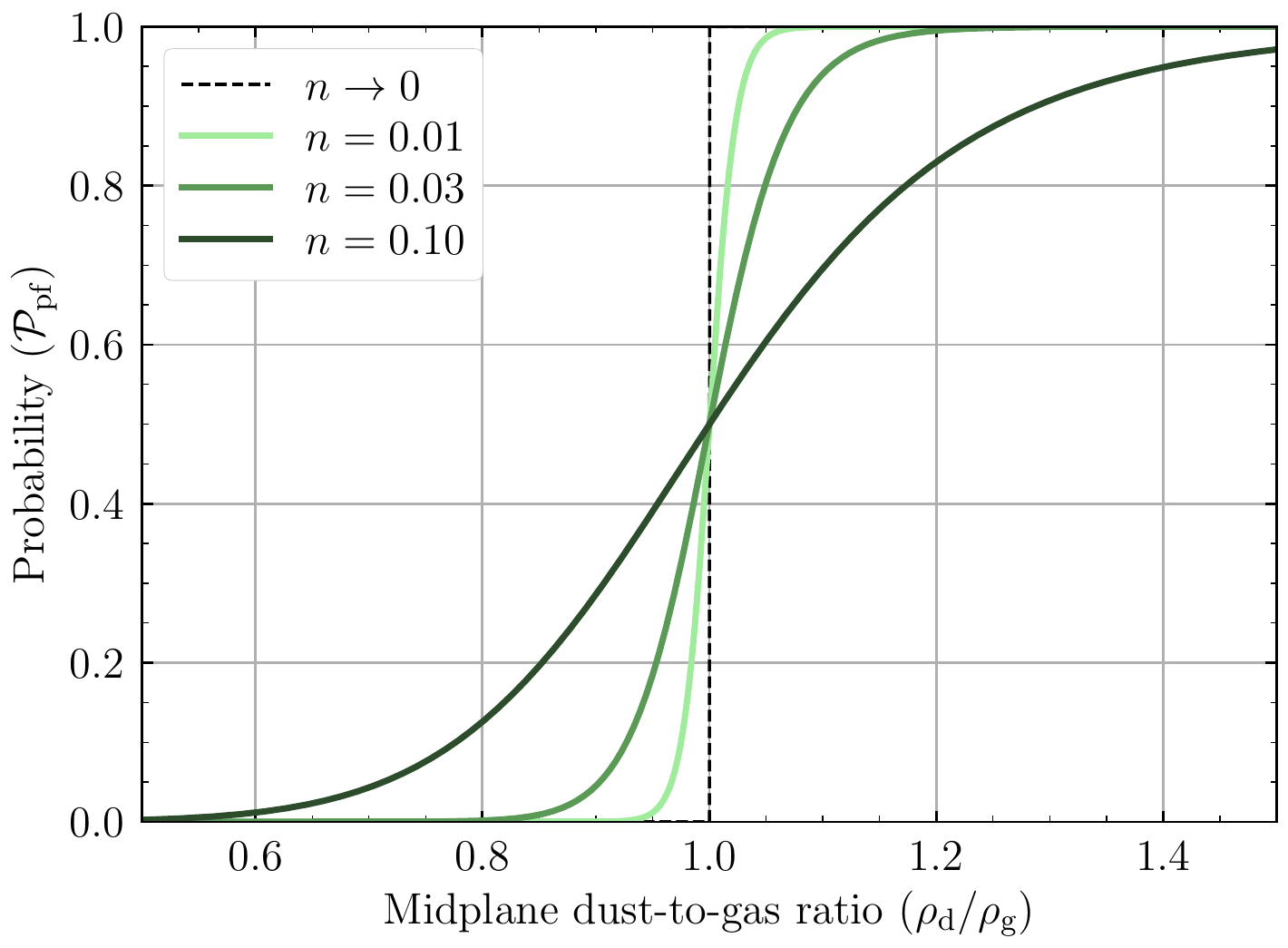}
    \caption{Planetesimal formation probability functions with varying smoothness values $n$ as a function of midplane dust-to-gas ratio.}
    \label{fig:pf_prob}
\end{figure}

The recipe used to transform dust mass into planetesimals is based upon the methods in \cite{Drazkowska2016} and \cite{Schoonenberg_2018}, which is outlined as follows. When the midplane dust-to-gas ratio exceeds 1, a fraction $\zeta=0.1$ of the dust mass per settling time scale of each mass bin $i$ is transformed into planetesimals as
\begin{equation}
    R^i = \frac{\partial}{\partial t} \Sigma^i_\text{d} = -\zeta \frac{\Sigma^i_\text{d}}{t_\text{sett}^i} = \mathcal{P}_{\rm pf}(-\zeta \Sigma^i_\text{d} \text{St}^i \Omega_\text{K}).
    \label{eq:pfrate}
\end{equation}
This trigger value of 1 is very conservative for the Stokes numbers that particles reach in our simulations at the dust traps ($\gtrsim0.1$). Recent simulations by \cite{Li2021} have shown that for Stokes numbers of $\sim0.1$, strong particle clumping occurs for midplane dust-to-gas ratios as low as 0.4. In addition, this threshold might not be an infinitely sharp transition and midplane dust-to-gas ratios slightly below the threshold could lead to some planetesimal formation. To imitate this more probable formation scenario, we smooth out the step-function centered at a dust-to-gas ratio of 1, i.e. $ \mathcal{P}_{\rm pf} = [0, 1]$, with a hyperbolic tangent planetesimal formation probability function, given by
\begin{equation}
    \mathcal{P}_{\rm pf} = \frac{1}{2}\Bigg[1 + \tanh{\Bigg(\frac{\log{(\rho_{\rm d}/\rho_{\rm g}})}{n}}\Bigg)\Bigg].
    \label{eq:pf_prob}
\end{equation}
We define $n$ as the smoothness parameter and note that this equation models the step function for $n \to 0$. Figure~\ref{fig:pf_prob} illustrates the difference of these methods. In our main simulations we take a smoothness value of $n = 0.03$, but in \S\ref{sec:sharp} explore the impact of other smoothness values. We also examine the effect of varying the planetesimal formation efficiency $\zeta$ in \S\ref{sec:zeta}. 

The summed rate of dust loss over all mass bins $i$ is then added to the planetesimal surface density
\begin{equation}
    \frac{\partial}{\partial t} \Sigma_{\rm plan} = -\sum_i R^i.
\end{equation}

\subsection{Creating the dust trap}
\label{sec:peturb}
Planetesimal formation requires high midplane dust-to-gas ratios, usually not achieved in smooth discs \citep[e.g.][]{garate}. The existence of a gap in the gas surface density profile is a mechanism to increase this midplane dust-to-gas ratio \citep[e.g.][]{Pinilla2012, Dipierro2015}, as particles can become trapped in the outer edge of this gap. This region just exterior to the gap is called the dust trap \citep{whipple_1972}. 
To model the effect of a dust trap, we create a bump in the viscosity parameter $\alpha$, which in turn induces a gap in the gas surface density. To do this, we first define a Gaussian profile following \cite{Dullemond_2018} with amplitude $A$, gap position $r_\text{g}$ and width $\omega$,
\begin{equation}
    F(r) =  \exp{\Bigg(\log{(A)}\exp{\Bigg(-\frac{(r-r_\text{g})^2}{2\omega^2}\Bigg)}\Bigg)}.
    \label{eq:bump}
\end{equation}
A bump in the viscosity profile is then achieved with
\begin{equation}
    \alpha(r) = \frac{\alpha_0}{F(r)},
    \label{eq:alpha}
\end{equation}
where $\alpha_0$ is the background $\alpha$ value, either $10^{-3}$ or $10^{-4}$. 

Before simulating a migrating gas gap, we first investigated the characteristics of a stationary gap that were more likely to produce planetesimals. In this stationary scenario, we explored the gap amplitudes $A = 3, 10$ and positions $r_{\rm g}$ = 30, 60 and 90 au. The width of the gap was not varied and simply set as the gas pressure scale height $H$. This is the ratio of the isothermal sound speed to the Keplerian frequency
\begin{equation}
   \omega = H = \frac{c_\text{s}}{\Omega_\text{K}},
\end{equation}
where the isothermal sound speed is defined as 
\begin{equation}
   c_{\rm s}^2 = \frac{k_{\rm B} T(r)}{\mu m_{\rm p}}.
   \label{eq:cs}
\end{equation}
Here $k_{\rm B}$ is the Boltzmann constant and $m_{\rm p}$ is the proton mass. We note that the gap width could impact many aspects of the model, such as the radial particle diffusion and the mass reservoir interior to the gap. As such, we explore different width values in \S\ref{sec:width} for robustness.

We chose to evolve a model with a pre-existing gap to eliminate evolutionary variability. As such, we also had to perturb the initial gas surface density by the Gaussian gap and renormalize the gas and dust surface densities for the initialisation.


Given this gap parametrisation, in our simulations we find that the dust trap is located at a distance that is 20\% and 30\% further than the gap for $A=3$ and 10, respectively. For a surface density proportional to $r^{-1}$, this relative distance scales with radius approximately\footnote{This exponent was found numerically by solving $\mathrm{d} P/\mathrm{d} r =0$.} as $r^{0.22}$. However, the surface density beyond 60~au is steeper than $r^{-1}$ given our initial conditions and slow viscous evolution, resulting in an almost constant relative distance between the gaps and traps.


\subsection{Dust trap migration}
\label{sec:migration}
For simplicity, we parametrise the radial velocity of the gap as a function of the velocity of gas inflow in a steady state disc, that is 
\begin{align}
    v_{\rm nominal} &= -\frac{3\nu}{2r} \\
    &= -\frac{3\alpha}{2} \frac{k_BT_\star}{\mu m_{\rm p}}  \Big(\frac{R_\star \sqrt{{\phi}}}{GM_\star}\Big)^\frac{1}{2} \nonumber \\
    &= -\alpha B. 
    \label{eq:vab}
\end{align}
This nominal velocity coincides with the standard velocity for planet under type-II migration \citep{Lin1986}.
Substituting in the disc initial conditions, we can express $B$ as a function of temperature at 50 au and star mass,
\begin{equation}
    B = 1.0 \times 10^{4}\ {\rm au~Myr}^{-1} \left(\frac{T_{\rm 50 au}}{40\ {\rm  K}}\right) \left(\frac{\ {\rm M}_\odot\ {\rm}}{M_\star }\right)^\frac{1}{2}.
    \label{eq:B}
\end{equation}
With $\alpha = 10^{-3}$ and $10^{-4}$ discs the nominal gap migration speeds are thus 10 and 1 au~Myr$^{-1}$, signifying over a simulation time of 10 Myrs the gap could migrate 100 and 10 au respectively. 

Since there are multiple scenarios in which a gap could migrate at a different speed \citep[see reviews by][]{Kley2012, Baruteau2014}, in our simulations for the migrating gap scenario we also explore other speeds expressed as fractions $f$ of this nominal velocity,
\begin{equation}
    v_{\rm gap} = f v_{\rm nominal} = -f\alpha B.
    \label{eq:fab}
\end{equation}
We use an initial gap position of 90 au because this would lead to a pressure maximum at about 100~au, which is close to the typical outer edge position of \textit{exoKuiper} belts around Solar-type stars \citep{Matra2018mmlaw}. In \S\ref{sec:wide} we explore the impact of having further out initial gap positions.

We test inward radial speeds at $f$ = 10, 30, 100 and 300\% of the nominal. Note that since the inner disc radius is 10 au, a gap travelling in an $\alpha = 10^{-3}$ disc at 10 au~Myr$^{-1}$ ($f$ = 100\%) will reach the simulated inner edge in 8 Myr.


\section{Results}
\label{sec:results}

The findings are separated into two sections:  \S\ref{sec:stat_results} describes the outcomes of various initial conditions for a stationary dust trap, and \S\ref{sec:moving_results} extends this investigation for the migrating dust trap. 

To compare the structure and content of the resulting planetesimal belts, we use four characteristics: width (\rw), center (\rc), fractional width (\rf) and planetesimal mass (measured in Earth masses M$_\oplus$). The width of each belt is defined as the distance upon which the planetesimal surface density remains $\geq 1\%$ of its maximum value. The inner and outer edges of in our simulated planetesimal belts are sharp, and so other choices of threshold minimally impact the measured width. The center is computed as the mean of the start and end points. 

\begin{figure*}
    \centering
    \includegraphics[width=\linewidth]{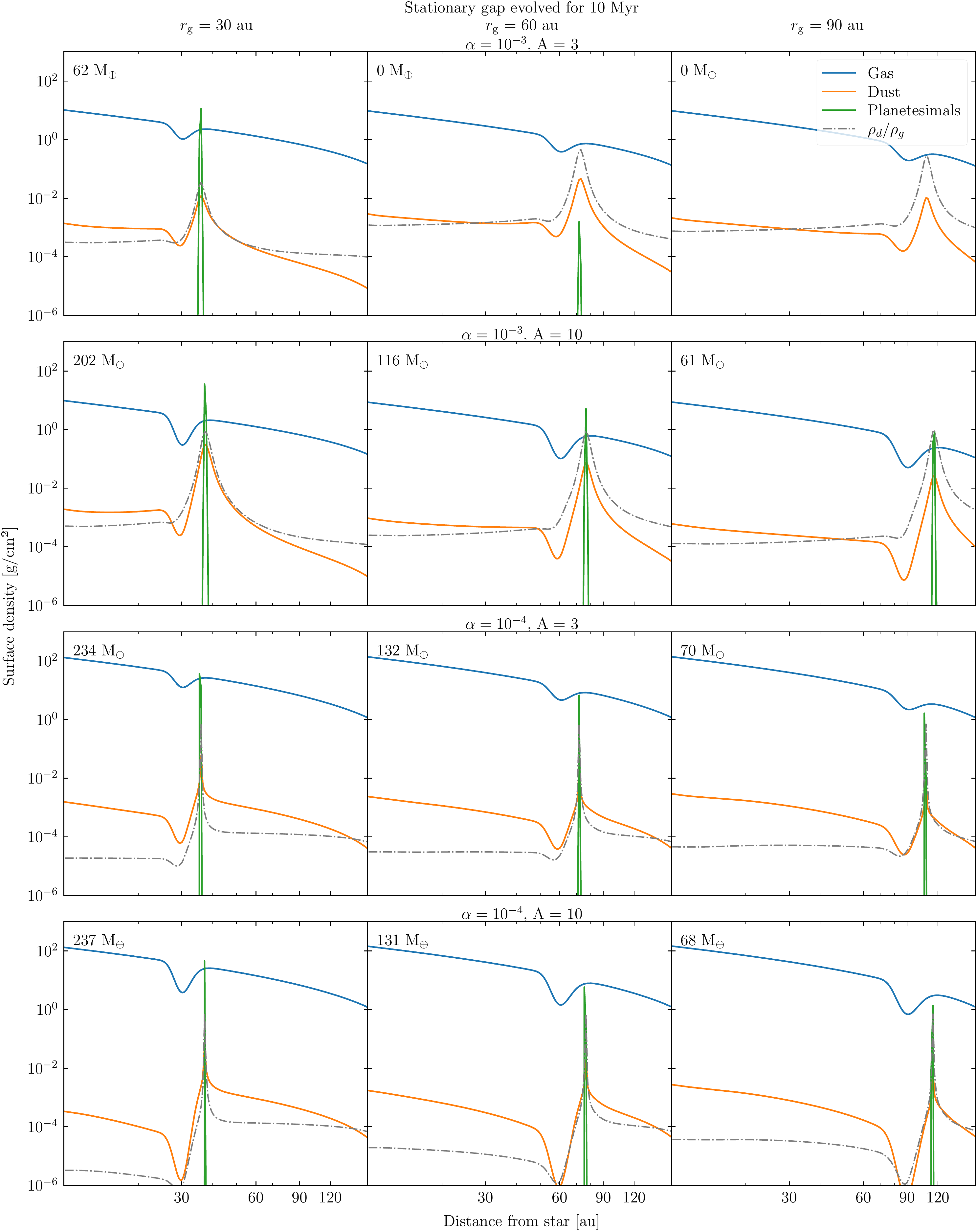}
    \caption{Gas, dust, and planetesimal surface density profiles of a protoplanetary disc with a stationary gap at $r_{\rm g}$ = 30, 60 and 90 au evolved for 10 Myr with varying viscosity parameter $\alpha$ and amplitude $A$. The dotted grey line displays the midplane dust-to-gas ratio $\rho_{\rm g}/\rho_{\rm d}$. The mass, width $\Delta r$ and fractional width $\Delta r/r$ of the resulting planetesimal belt are shown for each simulation.}
    \label{fig:sdr_stat_grid}
\end{figure*}

\begin{figure*}
    \centering
    \includegraphics[width=\textwidth]{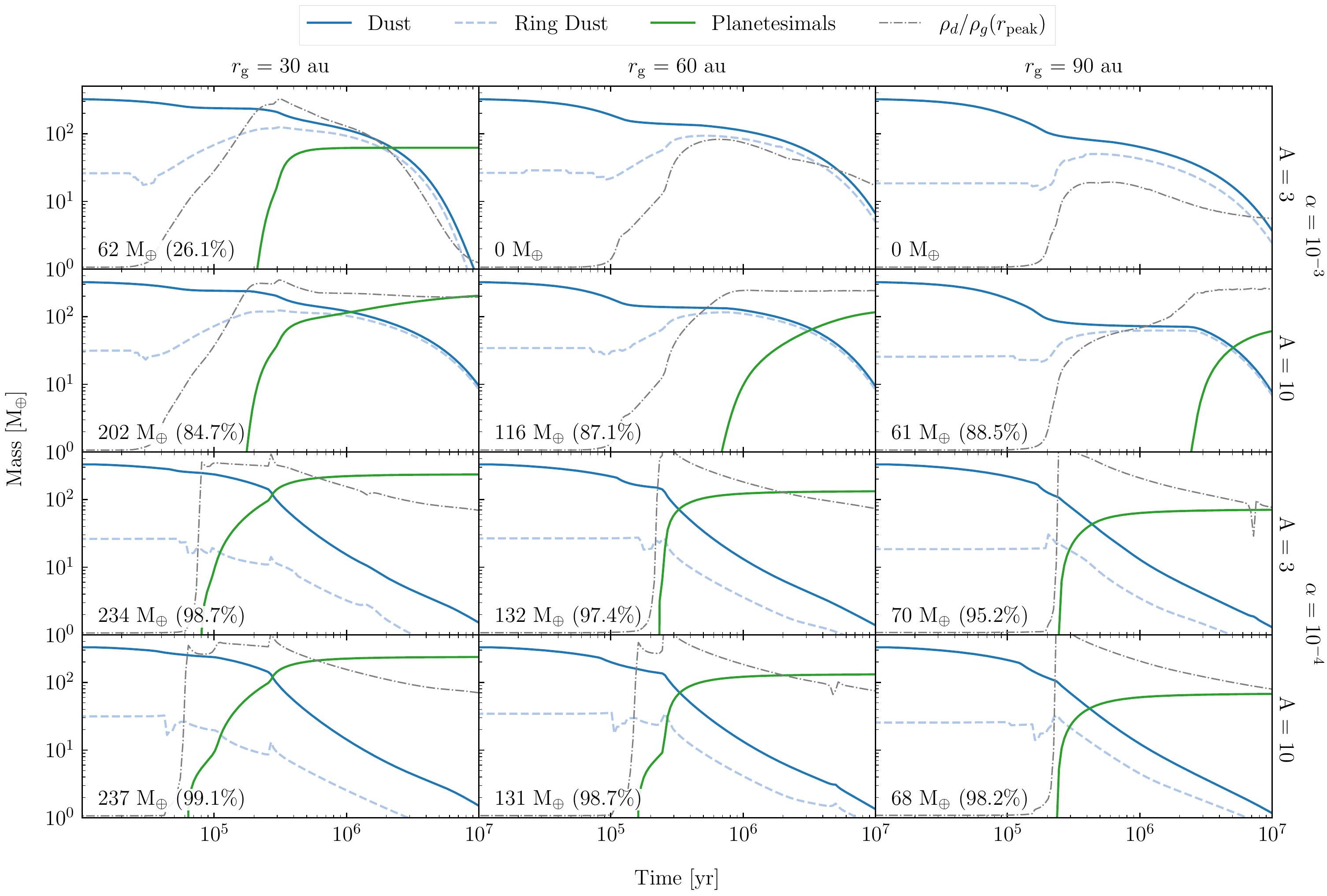}
    \caption{Mass evolution of the dust, ring dust and planetesimals in a protoplanetary disc with a stationary gap at positions $r_{\rm g}$ = 30, 60 and 90 au with varying viscosity parameter $\alpha$ and gap amplitude $A$. The final planetesimal mass for each simulation is presented, adjacent to the final percentage of the gap's initial exterior dust mass transformed into planetesimals. The dotted grey line displays the midplane dust-to-gas ratio at the dust peak $r_{\rm peak}$, which instead follows a linear scale from 0 to 1.}
    \label{fig:mass_stat_small}
\end{figure*}

\subsection{Stationary dust trap}
\label{sec:stat_results}

To determine the conditions favourable for planetesimal formation, we varied the viscosity parameter $\alpha$, gap amplitude $A$ and position $r_{\rm g}$. The final surface density profiles of the various simulations are show in Figure~\ref{fig:sdr_stat_grid}. The mass evolution and dust mass distribution for each case can be found in Appendix \S\ref{appendix_mass} and \S\ref{appendix_dist} respectively.

The two aspects of planetesimal formation we use to assess each simulation are the final efficiency and the timescale. We define the final formation efficiency as the percentage of initial exterior dust mass transformed to planetesimals by the end of the simulation. We define the initial exterior dust as all dust beyond $r_{\rm g}$ at $t = 0$. As dust grows it will drift inwards, and thus this solid mass reservoir roughly corresponds to the total mass available to form planetesimals. Note that this is only an approximation since a small fraction of the small dust interior to the gap will migrate out due to radial diffusion and viscous expansion and could contribute to planetesimal formation. To judge the formation timescale, we find the time taken for each simulation to form 1 M$_\oplus$ of planetesimal mass. This value was chosen as it coincides with the time the planetesimal formation rate is the highest, and thus is a good indicator of the formation timescale. By studying Figure~\ref{fig:mass_stat_small} (see Figure~\ref{fig:mass_stat_grid} for the full version), we deduce the conditions that favour each of these aspects are as follows.\\


\noindent \textit{PLANETESIMAL FORMATION FINAL EFFICIENCY}
\begin{enumerate}
    \item \textit{Low $\alpha$ discs.} \\
    In all simulation pairs with matching gap amplitude and position (e.g. $A=10$, $r_{\rm g} = 90$ au), $\alpha = 10^{-4}$ discs exhibit a higher formation efficiency (> 95\%) than $\alpha = 10^{-3}$ discs (e.g. compare rows 1 \& 3 in Figure~\ref{fig:mass_stat_small}). This is due to two effects: The lower $\alpha$ (and thus $\delta_{\rm r,z,t}$) decreases the radial diffusion and vertical stirring of dust; and also decreases the relative velocities of grains enabling their growth to larger sizes and Stokes numbers before fragmenting. Both effects combined lead to a stronger concentration of the dust mass near the pressure maximum at the gap edge and disc midplane. 
    \smallspace
    \item \textit{Gaps closer to the central star.} \\
    In simulation sets with matching $\alpha$ and amplitude, gaps at lower peak positions have a higher final planetesimal mass. This is due to gaps at smaller radii having larger exterior dust reservoirs available to be trapped and formed into planetesimals. The gap position has a minimal impact on the final planetesimal formation efficiency, however for the top right ($A = 3$, $\alpha = 10^{-3}$, $r_{\rm g} =$ 60, 90 au) simulations the final efficiency is 0\% since the dust-to-gas ratio never reaches values close to 1.
    \smallspace
    \item \textit{Steeper gaps}. \\
    In simulation pairs with matching $\alpha$ and gap position, a steeper gap amplitude leads to a higher final efficiency. We can explain this result with the fact that steeper gaps concentrate the dust in a narrower region, which helps reach the threshold dust-to-gas ratio more easily. Note that for the 60 and 90 au gaps in an $\alpha = 10^{-4}$ disc, the final planetesimal mass actually reduces slightly because the steeper gap profile makes less exterior dust available to the dust trap from the start.
\end{enumerate}

\noindent \textit{PLANETESIMAL FORMATION TIMESCALE}

\begin{enumerate}
    \item \textit{Low $\alpha$ discs.} \\ 
    At all gap positions, lower $\alpha$ discs produce planetesimals significantly faster. For example, the $r_{\rm g}$ = 90 au, $A = 10$ gap in an $\alpha = 10^{-4}$ disc produces 1 M$_\oplus$ of planetesimals an order of magnitude times faster than in an $\alpha = 10^{-3}$ disc ($\sim$ 0.2 Myr vs 2 Myr). See the last column in Figure~\ref{fig:mass_stat_small}. This is because the lower $\alpha$ leads to larger grains due to lower relative velocities, which drift faster and are more readily trapped at the dust trap and disc midplane. Therefore, the lower $\alpha$ is the easier is to concentrate dust and trigger the streaming instability.
    \smallspace
    \item \textit{Gaps closer to the central star}. \\
   Gaps closer to the star also produce planetesimals more quickly. For example, in the $\alpha = 10^{-3}$, $A = 10$ scenario forming 1 M$_\oplus$ of planetesimals takes $\sim$ 0.2, 0.7 and 2 Myr for gaps at 30, 60 and 90 au respectively (Figure~\ref{fig:mass_stat_small}, row 2). This is due to dust growth and dynamical timescales becoming shorter closer to the star as orbital speeds and relative velocities increase as well.
\end{enumerate}
Once the gap amplitude is sufficient to trigger planetesimal formation, increasing the amplitude further has a minimal impact on the time required to form 1 M$_\oplus$. In summary, we find low $\alpha$ discs and gaps closer to the central star are superior at forming planetesimals in both efficiency and timescale. We also find that if there is sufficient initial exterior dust available then increasing the gap amplitude results in more planetesimal mass, but does not significantly impact the formation timescale. 

\subsection{Migrating dust trap}
\label{sec:moving_results}

We studied the same scenarios ($\alpha = \{10^{-3},10^{-4}\}$, $A = \{3, 10\}$) for a gap initially at 90 au, travelling at 10, 30, 100 and 300\% of the nominal speed. We focused on this starting radius since the typical debris disc radius for a Solar-type star is $60-90$~au \citep{Matra2018mmlaw}, thus a starting gap position at 90~au could lead to a belt centre in that range. The final surface density profiles of the various simulations are show in in Figure~\ref{fig:sdr_mov_grid}. The mass evolution and dust mass distribution for each case can be found in Appendix \S\ref{appendix_mass} and \S\ref{appendix_dist} respectively.

The most notable finding of these simulations is that if the initial conditions were favourable to planetesimal formation with a stationary dust trap, dust can still effectively accumulate and form planetesimals as the trap migrates. The most striking example of this is an $A = 10$ gap travelling at the nominal velocity in an $\alpha = 10^{-3}$ disc, as shown in the case in the second row and last column of Figure~\ref{fig:sdr_mov_grid}. When migrating at $f = 100\%$, the gap creates a planetesimal belt 76 au wide with a fractional width of 1.39 after 7.6 Myr. This is nearly double the median fractional width of belts observed in mature systems, and close to the widest discs. Henceforth, we shall refer to this simulation as the \textit{prime} case. 

As expected, if the initial disc and gap conditions were not favourable for planetesimal formation in the stationary case, we find that moving the gap did not subsequently result in planetesimal formation. For example, from Figure~\ref{fig:sdr_mov_grid} it is evident that planetesimals formed for all velocities in all cases except for the $\alpha = 10^{-3}$, $A = 3$ set, since these conditions did not produce planetesimals when the gap was stationary at 60 and 90 au. Similar to the  stationary results, steepening the gap increases the dust-to-planetesimal transformation efficiency, but can result in a reduced overall planetesimal mass. 

\begin{figure*}
    \centering
    \includegraphics[width=\linewidth]{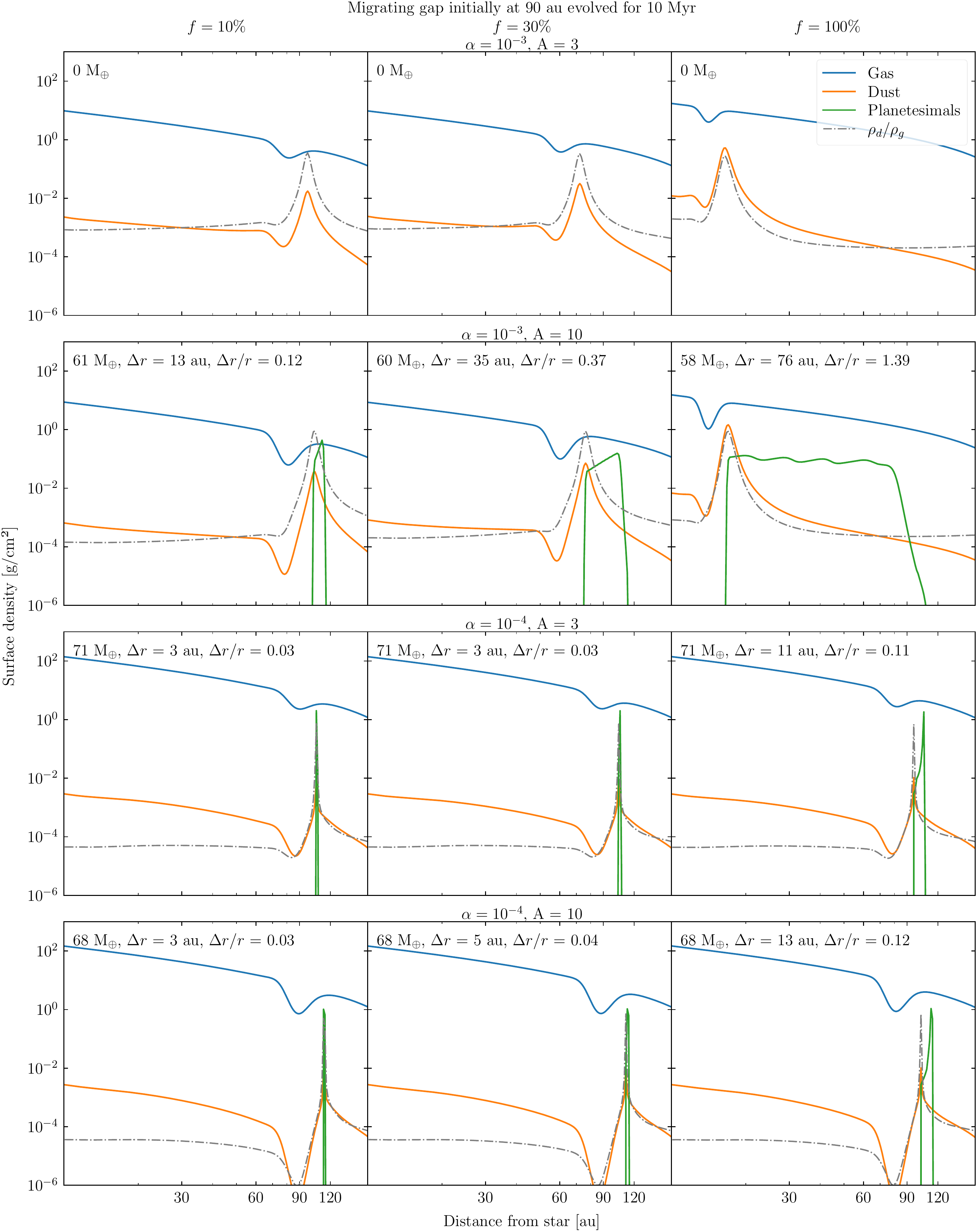}
    \caption{Gas, dust, and planetesimal surface density profiles of a protoplanetary disc with a gap initially at 90 au migrating at $f$ = 10, 30 and 100\% of the nominal velocity evolved for 10 Myr with varying viscosity parameter $\alpha$ and amplitude $A$. The dotted grey line displays the midplane dust-to-gas ratio $\rho_{\rm g}/\rho_{\rm d}$. The mass, width $\Delta r$ and fractional width $\Delta r/r$ of the resulting planetesimal belt are shown for each simulation. The snapshots presented for the two $\alpha = 10^{-3}$, $f = 100\%$ cases are instead at 7.6 Myr, before the gap reaches the simulation edge.}
    \label{fig:sdr_mov_grid}
\end{figure*}

\begin{figure}
    \centering
    \includegraphics[width=\columnwidth]{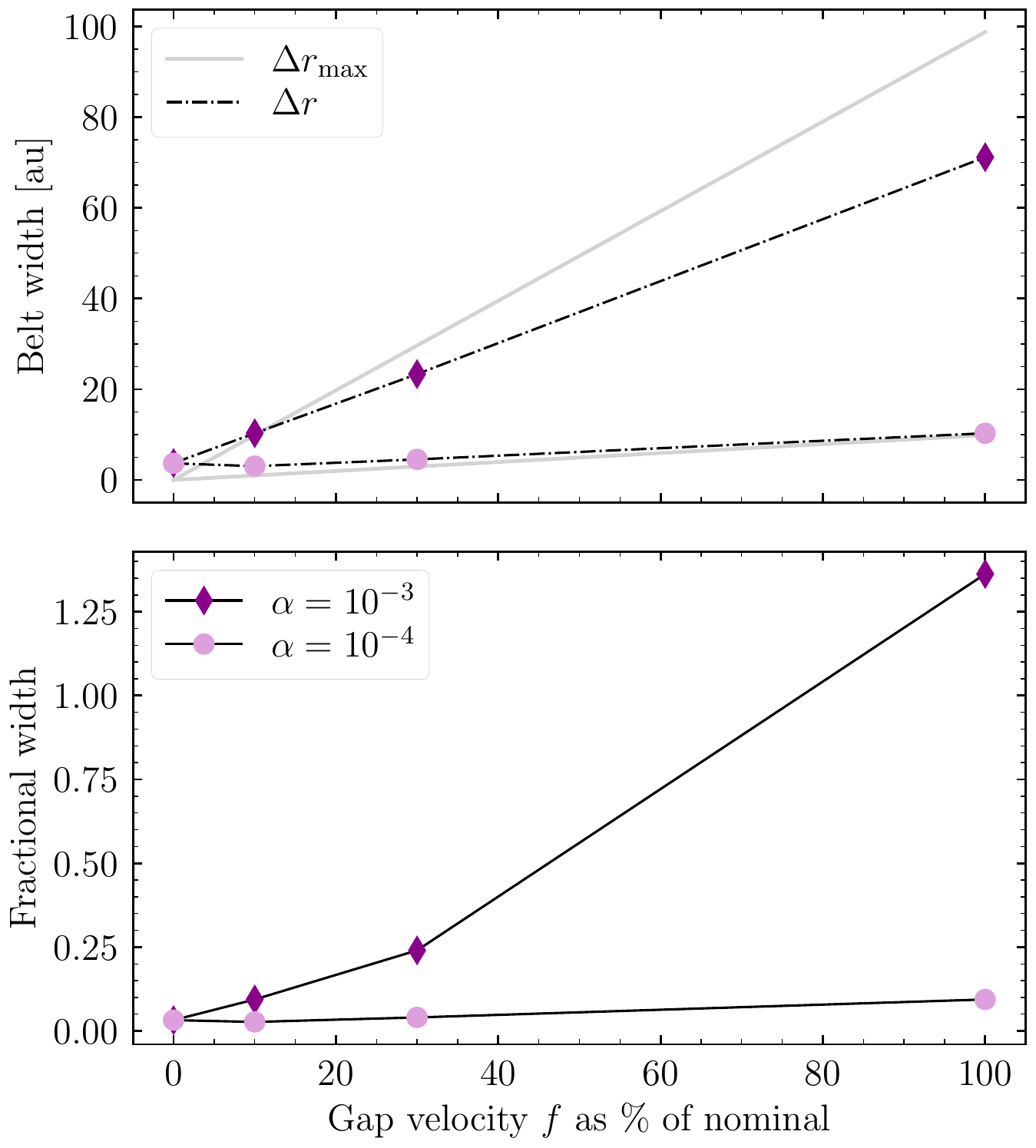}
    \caption{Simulated planetesimal belt widths and fractional widths as a function of gap migration velocity (for $A$ = 10). $\Delta r_{\rm max}$ is the hypothesised maximum width, defined in Equation~\ref{eq:max}. Note these are the statistics at $t = 7.6$ Myr, as the gap reaches the inner edge of the simulation just after this time for the $\alpha = 10^{-3}$ and $f$ = 100\% case.}
    \label{fig:ringstats}
\end{figure}

\begin{figure}
    \centering
    \includegraphics[width=\columnwidth]{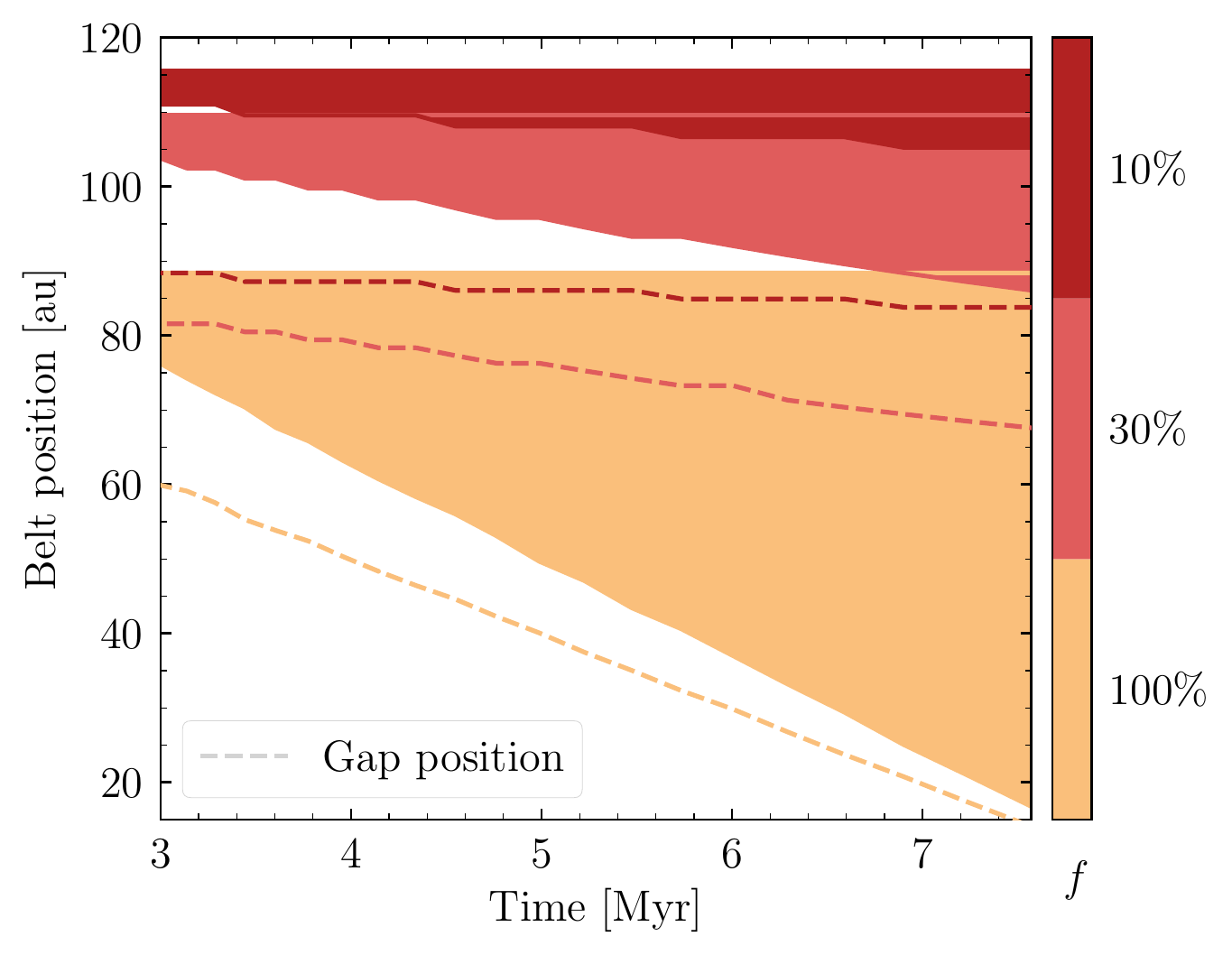}
    \caption{Belt and gap positions for the $A = 10$ gap in an $\alpha = 10^{-3}$ disc travelling at various percentages $f$ of the nominal velocity (colour coded on the right). The filled-in areas depict the disc regions over which planetesimals have formed.}
    \label{fig:belt_positions}
\end{figure}

Since dust can still effectively accumulate as the dust traps move, we observe the following qualitative correlations for both $\alpha$ discs. 
As the gap velocity increases, we find the planetesimal belt:
\begin{enumerate}
    \item Width increases
    \item Center position decreases
    \item Fractional width increases
\end{enumerate}
To fully appreciate these relationships and the quantitative impact of velocity, we need to study Figures~\ref{fig:sdr_mov_grid}, \ref{fig:ringstats} and \ref{fig:belt_positions} together. Note that once the gap is steep enough to form planetesimals, increasing the amplitude further only minimally impacts the resulting belt width.


Since the dust trap location is approximately at a distance that is $\sim$30\% further than the $A=10$ gap (see \S\ref{sec:peturb} for analysis,  Figure~\ref{fig:belt_positions} for results), the speed at which the dust trap moves and the distance it travels is 30\% larger as well, i.e.
\begin{equation}
    v_{\rm trap} \approx 1.3 v_{\rm gap}.
\end{equation}

Similar to the stationary results, migrating gaps in $\alpha = 10^{-4}$ discs form 1 M$_\oplus$ of planetesimal by 0.3 Myr (see Figure~\ref{fig:mass_mov_grid}, bottom two rows). By this time, gaps travelling at 10\%, 30\% and 100\% of 1 au~Myr$^{-1}$ only migrate $\sim$ 0.03, 0.1 and 0.3 au inwards, and the dust trap migrates $\sim$ 0.04, 0.13 and 0.4 au respectively. These distances are very small, and as such the outer edge of the planetesimal belt width is effectively constant against gap velocity. 

As the dust trap migrates further planetesimals continue to form, leading to an almost linear relationship between trap velocity and belt width. Since the width is small compared to the outer edge, we similarly obtain an almost linear relationship between trap velocity and fractional width for $\alpha = 10^{-4}$ discs, as shown in Figure~\ref{fig:ringstats}. For these cases with $A=10$, we can approximate the resulting planetesimal belt width as 
\begin{equation}
    \Delta r_{\rm max} \approx  v_{\rm trap} t \approx 1.3f\alpha B t .
    \label{eq:max}
\end{equation}

This estimation is also shown in Figure~\ref{fig:ringstats}, to compare with the actual simulated widths. It is evident that this estimation does not accurately approximate planetesimal belts produced in $\alpha = 10^{-3}$ discs, because the belt outer edge position decreases with increasing gap velocity (Figure~\ref{fig:belt_positions}). Similar to the stationary cases, migrating gaps in these  $\alpha = 10^{-3}$ discs take 3 Myr to form 1 M$_\oplus$ of planetesimal mass (see Figure~\ref{fig:mass_mov_grid}, top two rows). By this time, gaps travelling at 10\%, 30\% and 100\% of 10 au~Myr$^{-1}$ now migrate $\sim$ 3, 10 and 30 au inwards. Due to this, Equation~\ref{eq:max} becomes an upper limit of the belt width,
\begin{equation}
    \Delta r \leq 1.3 f\alpha B t.
    \label{eq:ineq}
\end{equation}

We also tested velocities at 300\%, and observe similar behaviour. The belt width becomes slightly larger in the $\alpha = 10^{-4}$ disc, but the gap in the $\alpha = 10^{-3}$ disc travels > 60 au before forming 1 M$_\oplus$ of planetesimal mass, leading to a much narrower belt.

\section{Discussion}
\label{sec:discussion}
We first analyse the planetesimal belt surface density profile magnitude and shape, evaluating the results against previous literature. We briefly discuss how our results could constrain the values of $\alpha$ at tens of au, then explore how other input parameters ($v_{\rm f}, \delta_{\rm r,z,t}, n, \zeta$ from Table~\ref{tab:ic}) impact the planetesimal  formation. Finally, we consider exotic scenarios in an attempt to widen the formed belt even further, and discuss our results in the context of current observations.

\subsection{Planetesimal belt surface density profile}

The planetesimal surface density profile in the prime case (Figure~\ref{fig:sdr_mov_grid} second row, last column) is relatively smooth, except for small ripples repeating over 10s of au. We find that this pattern is a numerical artefact of the simulation, rather than a physical phenomena; the amplitude of the ripples is reduced when increasing the number of radial bins or its resolution. The chosen radial resolution is found to be a good compromise between the running time and minimising the amplitude of the ripples. We also find that the ripple amplitude decreased  by increasing planetesimal formation probability smoothness $n$, and so higher resolutions are needed for smaller $n$ (see~\S\ref{sec:sharp}). Overall, we find the higher resolution simulations retain the belt width and approximate surface density, which will thus be the focus of this discussion.

\cite{Shibaike2020} estimated the planetesimal surface density from a migrating dust trap, using the assumption that once planetesimals start to form there is a quasi-steady state between the inward pebble  mass flux $\dot{M}_{\rm peb}$ and the formation of planetesimals $\dot{M}_{\rm plan}$. Here we aim to test this using as a reference epoch 5 Myr, when the dust trap is located at 50~au (see inner edge of Figure~\ref{fig:belt_positions}). At this time, the inward pebble mass flux just exterior to the trap (and relative to gap motion) was calculated to be $\sim 10^{-5}$ M$_\oplus$/yr. Using this value, the planetesimal surface density would be predicted to be \citep[Equation 21][]{Shibaike2020}
\begin{equation}
    \Sigma_{\rm plan} = \frac{\dot{M}_{\rm peb}}{2\pi r v_{\rm gap}} \approx 0.1\ \text{g~cm$^{-2}$} \left(\frac{\dot{M}_{\rm peb}}{10^{-5} \text{ M$_\oplus$/yr}}\right) \left(\frac{50\text{ au}}{r}\right) \left(\frac{10 \text{ au/yr}}{v_{\rm gap}}\right).
    \label{eq:shibaike}
\end{equation}
This value matches well with the surface density of planetesimals at 50~au in our simulations that is approximately 0.1~g~cm$^{-2}$. Since we find that the inward pebble mass flux exterior to the trap stays roughly constant after 1 Myr, if Equation \ref{eq:shibaike} holds the planetesimal surface density should be inversely proportional to $r$. However, as depicted in Figure~\ref{fig:slopes} in the appendix, we find a planetesimal surface density that is flatter with a more gentle negative slope that is indicative of $\dot{M}_{\rm plan}$ decreasing with time and becoming lower than the pebble flux.  This could be due to a limit on the planetesimal formation rate, in which case we would expect that the dust mass in the ring should increase as $\dot{M}_{\rm plan}<\dot{M}_{\rm peb}$. However, we find the opposite as the dust mass in the ring is decreasing with time after 3~Myr (see Figure \ref{fig:mass_mov_grid}). We interpret this dust mass loss as due to radial diffusion and drift across the gap. Therefore, we find $\dot{M}_{\rm peb}$ is not in equilibrium with $\dot{M}_{\rm plan}$. A weaker viscosity, dust radial diffusion, or a deeper gap could reduce the dust mass loss due to inward drift and thus produce a steeper planetesimal surface density profile. 


We would like to note as well that our finding of an approximately constant inward pebble mass flux of $10^{-5}$ M$_\oplus$~yr$^{-1}$ from 1 to 8 Myr is in contradiction to what we would obtain using the pebble predictor by \cite{Dr_kowska_2021}. With the pebble predictor, the pebble flux at the dust trap should decrease by two orders of magnitude between 1-10 Myr. This difference is likely due to the initial viscous expansion of the disc in our \textsc{DustPy} simulations and the presence of the gap.

\subsection{Constraining $\alpha$}
The inequality of Equation~\ref{eq:ineq} derived from our simulations could potentially be extended to infer the minimum $\alpha$ value in protoplanetary discs if the width of debris discs is indeed a result of this migration. If we assume that a gap cannot travel faster than the nominal velocity ($f = 1$), e.g. as in type-II planet migration, then the width of debris discs  ($\Delta r$) could constrain the allowable values of $\alpha$ in their progenitor protoplanetary discs to
\begin{equation}
    \alpha \gtrsim \frac{\Delta r}{1.3 Bt}.
\end{equation}


As predicted in \S\ref{sec:migration}, migrating dust traps in $\alpha = 10^{-4}$ discs move too slowly to form wide planetesimal belts due to the proportionality between $\alpha$ and gap velocity (see Figure~\ref{fig:sdr_mov_grid}). If wide planetesimal belts are indeed formed by migrating dust traps, then the large widths of observed belts (median value of 50~au for FGK stars) constrain $\alpha$ to values 

\begin{equation}
    \alpha\gtrsim 4\times10^{-4} \left(\frac{\Delta r}{50\ {\rm au}}\right) \left(\frac{10\ {\rm Myr}}{t_{\rm migration}}\right) 
     \left(\frac{5772\ {\rm K}}{T_\star}\right)  \left(\frac{2\ {\rm R}_\odot\ {\rm}}{R_\star }\right)^\frac{1}{2}\left(\frac{M_\star }{\ {\rm M}_\odot\ {\rm}}\right)^\frac{1}{2}.
\end{equation}
It is important to note here that the derivation above is for the case of one dust trap only. It is possible that multiple rings in low $\alpha$ regions (travelling more slowly) could also form wide planetesimal belts. Protoplanetary discs with multiple rings are in fact common \citep{Andrews2018, Long2018, Cieza2021}, and a few well resolved wide \textit{exoKuiper} belts with ALMA show gaps \citep{Marino2018, Marino2019, Marino2020, MacGregor2019, Daley2019, Nederlander2021}. However, simulating multiple gaps would require further assumptions on to what happens to planetesimals that are swept by a gap \citep[e.g. these could be scattered or accreted by the gap-forming planet,][]{Fernandez1984, Ida2000, Gomes2004, Kirsh2009, Eriksson2020, Eriksson2021}, and thus studying the effect of multiple gaps is beyond the scope of the paper.


\subsection{Reduced fragmentation velocity and $\delta$}
\label{sec:vfrag}
As pointed out in \S\ref{sec:fragmentation}, the fragmentation velocity is an uncertain quantity which can have a great impact on the simulation outcomes. This is because the growth of dust trapped at the pressure maximum is limited by fragmentation (i.e. by the fragmentation barrier). Hence a different fragmentation velocity would change the maximum dust size (or Stokes number) and thus the efficiency of trapping. The results presented in previous sections used a fragmentation velocity of 10~m~s$^{-1}$, but here we aim to explore the impact of lower values, and show as well that a lower fragmentation velocity combined with a lower level of turbulence could still lead to the same results. This is because the maximum grain size in the fragmentation barrier is approximately \citep{Birnstiel_2010}
\begin{equation}
a_{\rm f}=\frac{\Sigma_{\rm g} v_{\rm f}^2}{\rho_{\rm s} \upi \delta_{\rm t} c_{\rm s}^2 }.
\label{eq:vf}
\end{equation}
Recent laboratory experiments that have studied the sticking properties of dust grains have suggested fragmentation velocities lower than 10~m~s$^{-1}$ \citep[e.g.][]{gundlach_2018, Musiolik_2019, Steinpilz_2019}. As such, we start by exploring fragmentation velocities of 1 and 3~m~s$^{-1}$. We investigate these conditions for the stationary dust trap set, as our results demonstrate that if formation occurs in this scenario then planetesimals are likely to also form as the trap migrates.  

For an $\alpha = 10^{-4}$ disc, we find similar results just with a lower final planetesimal mass. This suggests that the primary findings would still be valid for lower $v_{\rm f}$, but only at radii where planetesimal formation occurs (e.g. for $v_{\rm f}$ = 1~m~s$^{-1}$ for $r < 60$ au). This agrees with the findings of \cite{Pinilla_2021}. However, we find that planetesimals either do not form or produce mass an order of magnitude lower for a stationary dust trap in an $\alpha = 10^{-3}$ disc with fragmentation velocities of 1 or 3~m~s$^{-1}$. From Equation~\ref{eq:vf}, we can see that a change in $v_{\rm f}$ can be balanced by a change in $\delta_{\rm t}$. Following the work of \cite{Pinilla_2021}, we explore several scenarios where $\delta_{\rm r,z,t}\leq\alpha$. Refer to \S\ref{sec:dust} for a description of the individual $\delta$ parameters. In order to balance a factor 10 lower $v_{\rm f}$, we need to reduce $\delta_t$ by a factor 100 to achieve the same $a_{\rm f}$. The parameters explored for the $\alpha = 10^{-3}$, $A = 10$, $f = 0\%$ scenario (stationary counterpart of prime case) and results are outlined in Table~\ref{tab:delta}.

\begin{table}
    \centering
    \caption{Final planetesimal masses for the $\alpha = 10^{-3}$, $A = 10$, $f = 0\%$ scenario under varying fragmentation velocities $v_{\rm f}$ and $\delta_{\rm r,z,t}\leq\alpha$.}
    \begin{tabular}{llllc}
        \hline
         $\delta_{\rm t}$ & $\delta_{\rm r}$ & $\delta_{\rm z}$ & $v_{\rm f}$ (m~s$^{-1}$) & Planetesimal Mass (M$_\oplus$)\\
         \hline
         $10^{-3}$ &  $10^{-3}$ &  $10^{-3}$ & 1 & $\times$ \\
         & & & 3 & 6 \\
         & & & 10 & 58 \\
         \hline
         $10^{-5}$ &  $10^{-5}$ &  $10^{-5}$ & 1 & 71\\
         & & & 3 & 70\\
       \hline
          &  $10^{-3}$ &  $10^{-5}$ & 1 & 64 \\
         & & & 3 & 69\\
        \hline
          &  $10^{-5}$ &  $10^{-3}$ & 1 & 67 \\
         & & & 3 & 70\\
         \hline
    \end{tabular}
    \label{tab:delta}
\end{table}

From these findings, we see that although $\delta_{\rm r,z,t} = \alpha$ either forms no planetesimals or a planetesimal mass an order of magnitude lower at a reduced fragmentation velocity, lowering $\delta_{\rm t}$ and either $\delta_{\rm r}$ or $\delta_{\rm z}$ to compensate results once more in planetesimal formation. The exact values of $\delta_{\rm r}$ or $\delta_{\rm z}$ in protoplanetary discs with substructures is uncertain, but modelling of the vertical and radial thinness of substructures could help to constrain these values further. For example, \cite{Dullemond_2018} showed that the rings seen in DSHARP are inconsistent with $\delta_{\rm r}\ll5\times10^{-4}$ and grains sizes above 1 mm. This constraint would rule-out the models with $\alpha=10^{-4}$ and $\alpha=\delta_{\rm r,z,t}$ since in those models grains grow beyond cm sizes, but it does not rule-out the parameters explored in this section since grains do not grow beyond mm sizes. On the other hand, \cite{Pinte2016} and \cite{Villenave2020} found very effective vertical settling of mm-sized grains, which in the case of HL~Tau it corresponded to $\delta_{\rm z}<10^{-3}$. Therefore, the parameters explored here are consistent with the current observational constraints on $\delta_{\rm r,z}$. 


\subsection{Planetesimal probability function}
\label{sec:sharp}
\begin{figure}
    \centering
    \includegraphics[width=\columnwidth]{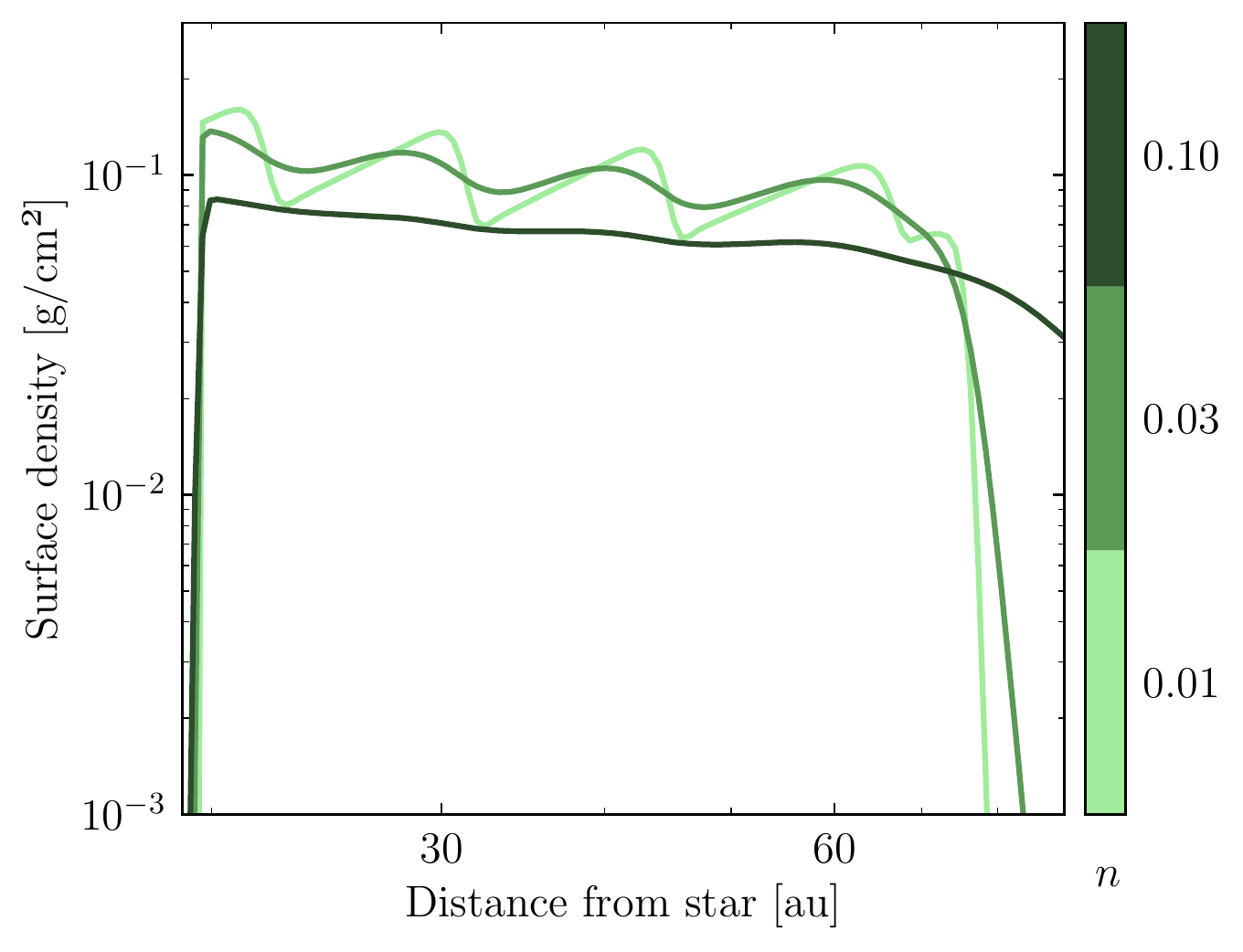}
    \caption{Impact of the planetesimal probability function smoothness $n$ (see Figure~\ref{fig:pf_prob}) on the planetesimal surface density profile for the prime case ($\alpha = 10^{-3}$, $A = 10$, $f = 100\%$). The original simulation used $n = 0.03$.}
    \label{fig:planrate}
\end{figure}

To check the robustness of our results, we explore the impact of modifying the midplane dust-to-gas ratio range that forms planetesimals. We run two more prime case simulations ($\alpha=10^{-3}$, $A = 10$, $f = 100\%$) with a varied probability smoothness $n$ value from Equation~\ref{eq:pf_prob}, and test both a wider and narrower formation range (see Figure~\ref{fig:pf_prob}). The results are shown in Figure~\ref{fig:planrate}.

As expected, the smoothed probability functions (with higher $n$) form planetesimals more readily due to the lowered midplane dust-to-gas threshold. This means that a wider formation range also leads a wider final planetesimal belt (up until the limit where planetesimals form at the initial dust trap position, as was reached by $n = 0.1$). Interestingly, we also find that a steeper probability function (as $n \to 0$) produces much sharper and higher amplitude ripples. This shows that the ripples are sensitive to both the grid resolution and the smoothness of the planetesimal formation probability function, i.e. steeper planetesimal formation functions require higher radial resolutions.

\begin{figure}
    \centering
    \includegraphics[width=\columnwidth]{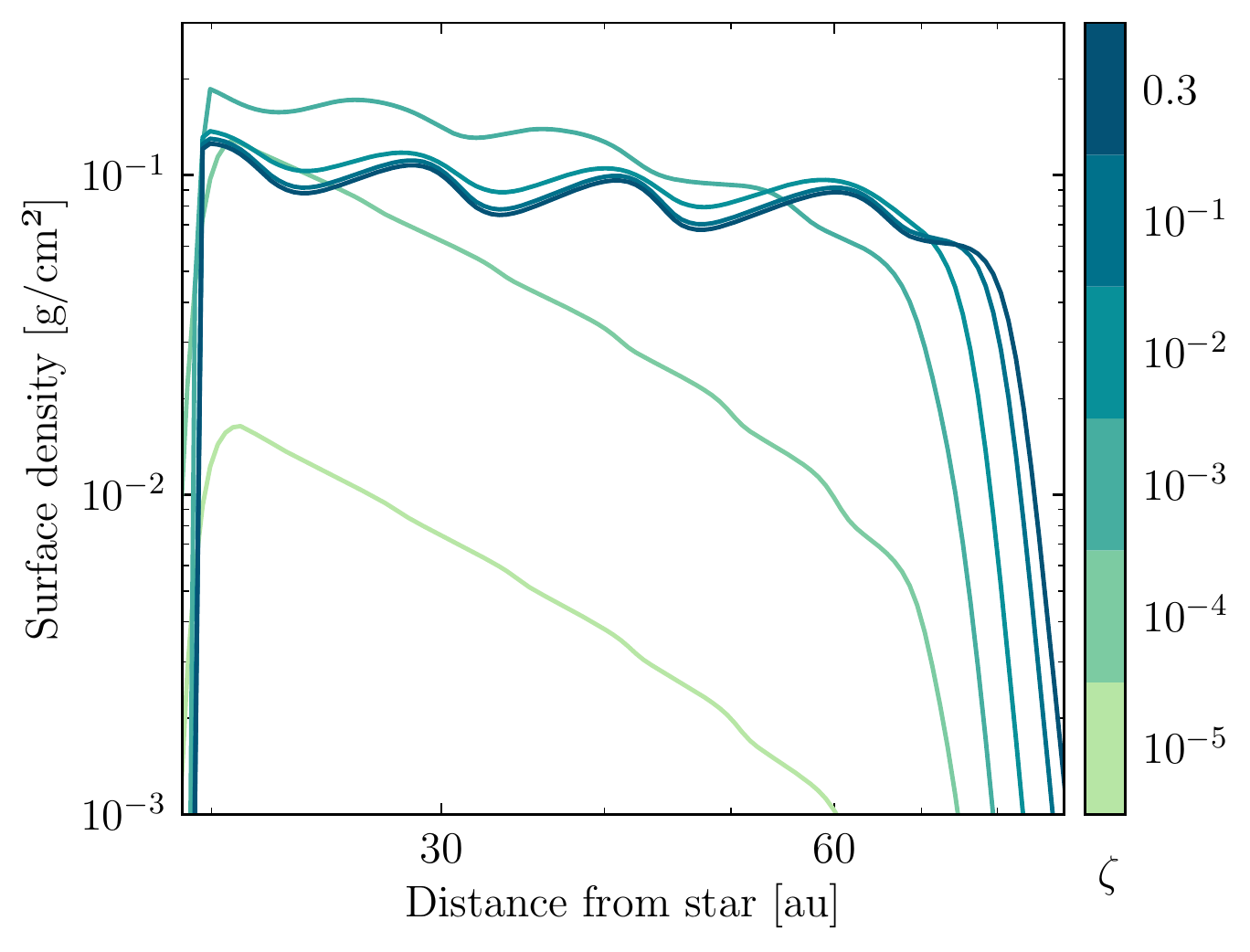}
    \caption{Impact of planetesimal formation efficiency $\zeta$ on the planetesimal surface density profile for the prime case ($\alpha = 10^{-3}$, $A = 10$, $f = 100\%$). The original simulation used $\zeta = 0.1$.}
    \label{fig:zeta}
\end{figure}

\subsection{Planetesimal formation efficiency}
\label{sec:zeta}
The planetesimal formation rate is proportional to the efficiency parameter $\zeta$ (Equation~\ref{eq:pfrate}), which determines the fraction of dust mass per settling timescale to transform into planetesimals.  In the main simulations we used $\zeta = 0.1$, however here we explore the impact of increasing and decreasing this parameter.

From Figure~\ref{fig:zeta}, we find that the disc's outer edge is particularly sensitive to $\zeta$. Higher efficiencies enable rapid planetesimal formation in the outer regions and thus the final planetesimal disc is more extended. However, we find that at a given radius the planetesimal formation rate saturates at a given $\zeta$ and increasing the efficiency does not lead to a higher surface density. This likely happens when the planetesimal formation rate becomes equal to the rate at which solids are being resupplied by the inward pebble flux. Conversely, lower values of $\zeta$ lower the planetesimal formation rate, which is not longer regulated by the inward pebble flux but rather $\zeta$ and the dust mass in the disc. We find that some lower efficiencies (e.g. $\zeta = 10^{-3}$) lead to an overall higher belt surface density interior to 60~au compared to higher $\zeta$ values. We interpret this as a result of mass conservation \citep{Lenz2019, Lenz_2020}. Pebbles that are not converted into planetesimals in the outer disc can drift inward or move with the trap and form planetesimals further in later on. In fact, we find that the total mass converted into planetesimals for $\zeta=0.3, 10^{-1}, 10^{-2}, 10^{-3}, 10^{-4}, 10^{-5}$ is 57, 57, 56, 52, 15 and 2 M$_\oplus$ respectively, demonstrating how similar the total planetesimal masses are for $\zeta>10^{-4}$.


\subsection{Gap width}
\label{sec:width}

\begin{figure}
    \centering
    \includegraphics[width=\columnwidth]{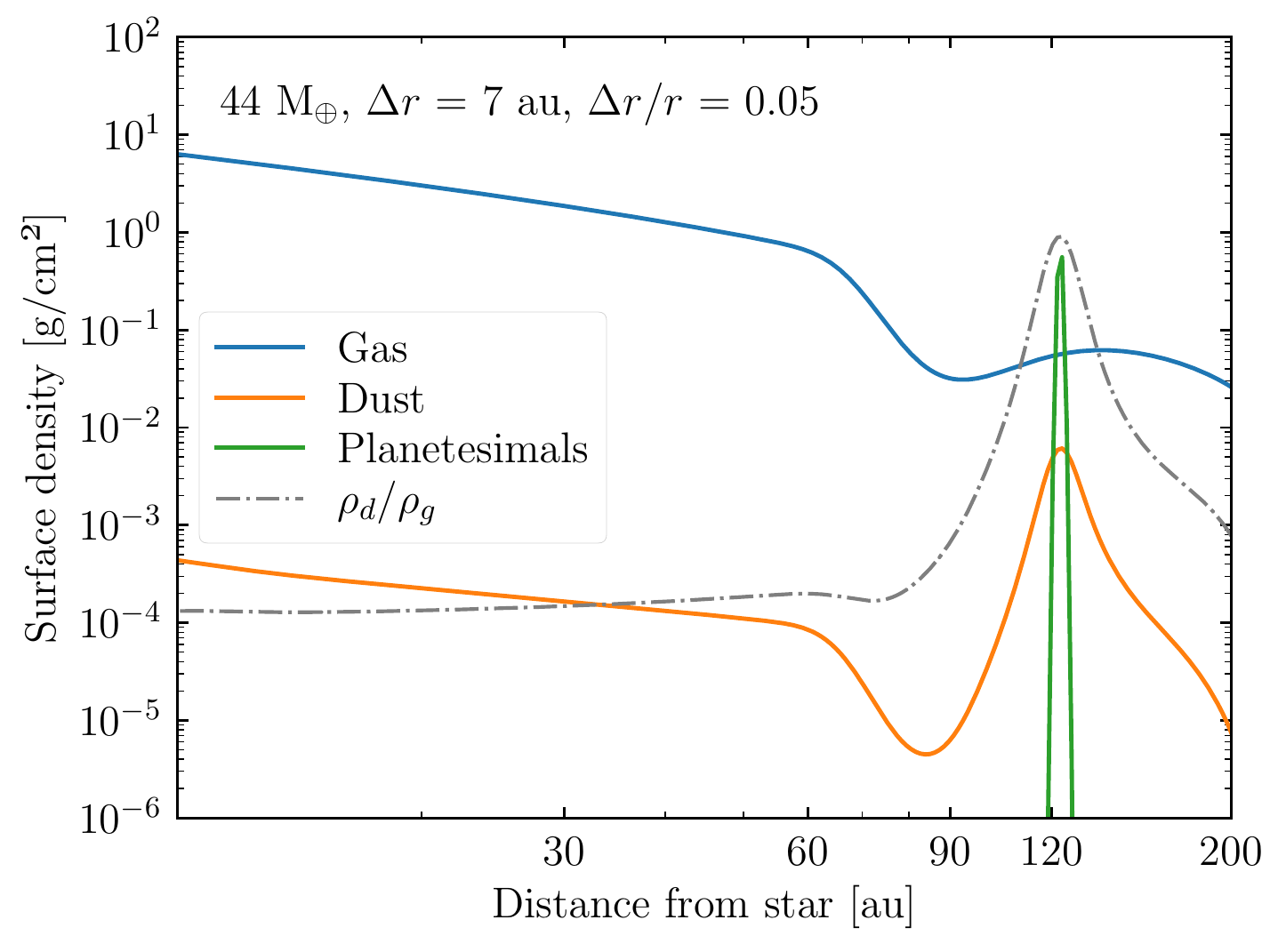}
    \caption{Stationary counterpart of prime case simulation ($\alpha = 10^{-3}$, $A = 10$, $f = 0\%$) with double the gap width, i.e. $\omega_{\rm gap}$ = $2H$ at 10 Myr.}
    \label{fig:width}
\end{figure}

As noted in \S\ref{sec:peturb} the width of the gap plays an important role in the dust evolution since, together with the gap amplitude, it affects how steep the local pressure maximum is and thus how efficient is the dust trapping. Hence, for robustness we explored the influence of a gap twice as wide (i.e. a width double the gas pressure scale height) on the stationary counterpart of the prime case simulation ($\alpha = 10^{-3}$, $A = 10$, $f = 0\%$). The result is shown in Figure~\ref{fig:width}, which demonstrates planetesimals can still successfully form under these conditions. Given that the original simulation with $\omega = H$ produced a 61 M$_\oplus$ planetesimal belt of width $\Delta r$ = 3.7 au, as expected doubling the gap width leads to a lower mass of produced planetesimals (44~M$_\oplus$), and a wider planetesimal belt (6.6 au) since dust grains are less concentrated around the pressure maximum. Note that while in this work we have treated the width and depth of the gas gap independently, these properties can be closely related if the gaps are caused, for example, by planets \citep{Lin1979}. Finally, increasing the width of the gap also shifts the trap location to a larger radius, which would lead to an even wider planetesimal belt if the gap was migrating.

\subsection{Extending the width of the belt}
\label{sec:wide}
We established in \S\ref{sec:moving_results} that fast-moving dust traps in $\alpha = 10^{-3}$ discs can travel tens of au before accumulating enough dust to form planetesimals. This behaviour limits the belt width, so here we experiment with producing even wider belts under similar initial conditions. To do so we devise three situations to establish whether this lack of initial formation is driven by insufficient time or dust at larger radii.

To observe the effect of time, we fixed the prime case gap to remain stationary at 90 au for 1 Myr before beginning its migration. To observe the effect of dust availability, we moved the initial gap position from 90 to 120 au and allowed it to migrate immediately. Finally, we combined these aspects together and simulated a gap initially at 120 au held stationary for 2 Myr before migrating. 

\begin{figure}
    \centering
    \includegraphics[width=\columnwidth]{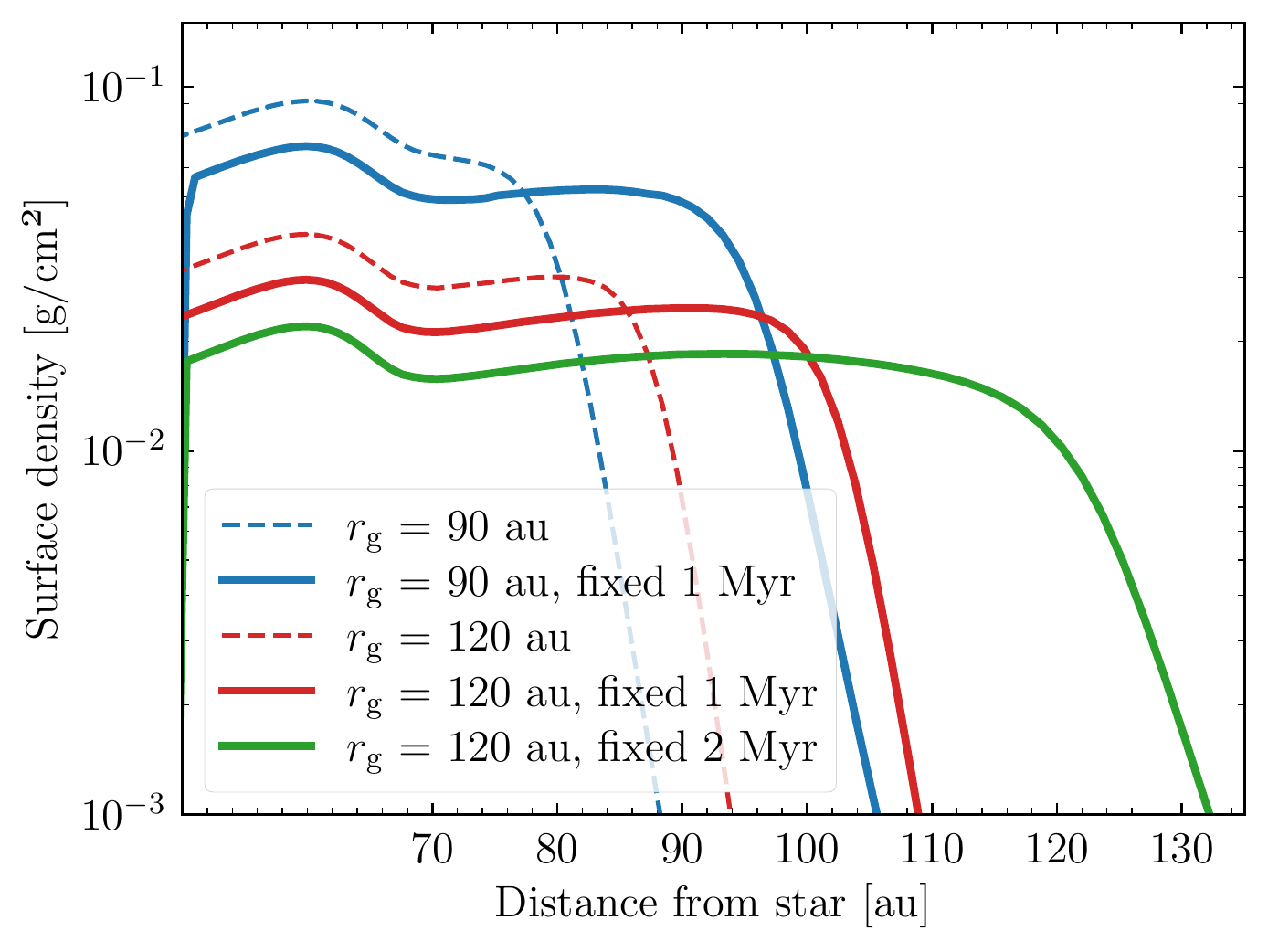}
    \caption{Impact of initial gap position and time the gap is initially fixed on the planetesimal surface density profile for the prime case ($\alpha = 10^{-3}$, $A = 10$, $f = 100\%$). The original simulation used $r_{\rm g} = 90$ au.}
    \label{fig:experimentation}
\end{figure}

By considering the results in Figure~\ref{fig:experimentation}, we can see that the widest planetesimal belts are produced by holding the migrating gap stationary, when compared to the counterpart simulations. This is due to the long timescales for dust growth and inward drift at those large radii, and it can be observed that planetesimals only form at $r > 90$ au when the gap is held stationary for 1-2 Myr. This is consistent with the results from Figure~\ref{fig:mass_stat_small}, which shows that planetesimal formation only begins at 2 Myr for a stationary $\alpha = 10^{-3}$, $A=10$, $r_{\rm g} = 90$~au  gap.

The surface density of the wider belts are slightly lower, but this is due to mass conservation as both simulations produced a very similar total planetesimal mass. This result suggests that insufficient time for the dust to accumulate \textit{is} the primary reason for the lack of  planetesimal formation between 80-100~au for fast gaps in $\alpha = 10^{-3}$ discs, rather than the lower dust availability further out in protoplanetary discs. Note that this need to wait for dust to evolve and accumulate in the pressure bump is consistent with observations. The observed protoplanetary discs with substructures between 30-230~au have a wide distribution of ages from less than 1~Myr to 10~Myr \citep{Andrews2018, Long2018, Cieza2021}. Therefore, it is reasonable to consider dust traps beyond 100~au in discs older than 2~Myr.

When we start the gap at 120~au instead of 90~au, we manage to significantly shift the outer edge of the belt when the gap is fixed for 2 Myr. This suggests that at larger radii, dust traps need to remain stationary for longer since the dust growth and drift timescales are larger. As previously mentioned, this is consistent with the results from Figure~\ref{fig:mass_stat_small}, which demonstrates planetesimals in this system take $>$2 Myr to form.

\subsection{Comparison with observations}
\label{sec:comparison}

\begin{figure}
    \centering
    \includegraphics[width=\columnwidth]{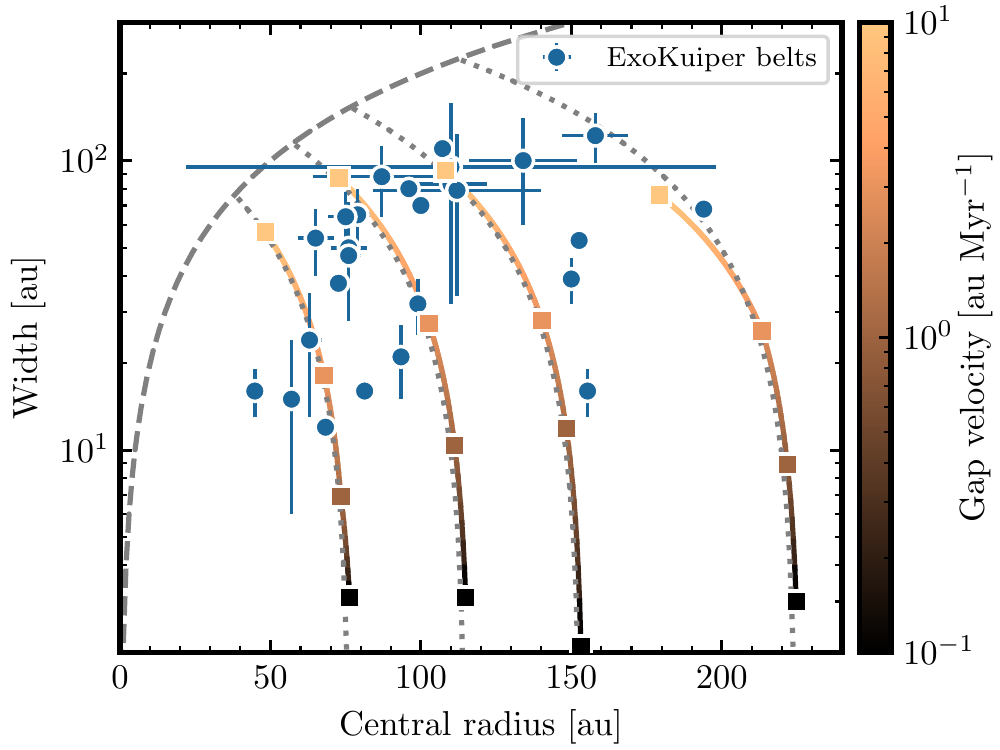}
    \caption{Width vs radius of \textit{exoKuiper} belts around FGK stars compared with the results from our simulations after 7.6 Myr. The blue errorbars represent the results for FGK stars ($L_{\star}=0.1-5.5$ L$_{\odot}$) of the REASONS survey (Matra et al. in prep), where the width is the full width half maximum (FWHM) of the belt assuming a Gaussian radial distribution. The square markers represent the results from $\alpha=10^{-3}$ simulations with gaps starting at 60, 90, 120 and 180~au. The colour bar represents the velocity of the gap, with the markers indicating $f = $ 0, 10, 30, and 100\% the nominal speed. The solid lines connecting these points is a linear interpolation. The grey dashed line simply displays the maximum width of a belt, that is twice its central radius. The grey dotted lines show the width as a function of radius for belts with a fixed outer edge position ($r_{\rm out}$), i.e. width=2($r_{\rm out}$-r)}.
    \label{fig:comp}
\end{figure}

Here we aim to put our findings into the context by comparing our simulation results with the width and radius of \textit{exoKuiper} belts around FGK stars (defined here as $L_{\star}=0.1-5.5$ L$_{\odot}$) as measured by the the ALMA REASONS survey (Matra et al. in prep). Figure~\ref{fig:comp} shows in blue the observed width and radius including $1\sigma$ uncertainties. This sample of \textit{exoKuiper} belts has a median central radius of 95~au, width of 54~au, and a fractional width of 0.63. The coloured squares show the width and radius of planetesimal belts as a function of gap velocity (0, 10, 30 and 100\% of the nominal speed), with gap positions of 60, 90, 120 and 180~au and $\alpha=10^{-3}$. Given our findings of \S\ref{sec:wide}, we kept the gaps stationary for 2~Myr for gaps starting at 60 and 90~au, and for 3~Myr for gaps starting at 120 and 180~au in order to be able to form planetesimals at large radii. This waiting period results in an outer edge position that is independent of the gap velocity (very close to the grey dotted lines that represent a fixed outer edge position). For the gaps starting at 120 and 180~au, we evolved them for 10~Myr, and for the gaps starting at 60 and 90~au, we evolved them for 6.3 and 8.6 Myr as they reach the inner boundary before 10~Myr.

This comparison shows that the spread of widths could be achieved by different gap velocities (or different periods over which a gap migrates), and the spread in radius by different initial locations where planetesimals started forming. Overall, we find that all these belts are wider than the width achieved by the gaps moving at 10\% the nominal speed (1~au~Myr$^{-1}$), and thus simulations with $\alpha=10^{-4}$ would not be capable of reproducing the large widths. Similarly, to reproduce the widest belts with widths of $\sim100$~au, gaps need to travel at 100\% of the nominal speed (10~au~Myr$^{-1}$) or even faster if discs live shorter timescales. In order to reproduce the median values for the radius and width of \textit{exoKuiper} belts, the gap should start at roughly 90~au and travel at $\sim10$~au~Myr$^{-1}$. Therefore, we conclude that the scenario presented in this paper can reproduce the main characteristics (radius and width) of exoKuiper belts.

Note that the apparent correlation between the widths and radii of observed exoKuiper belts is simply due to an upper limit on the width of a belt. Namely, a belt cannot have a width that is twice its central radius (dashed line). In fact, narrow belts are found in belts with both small and large central radius. Similarly, the anti correlation between the width and radius shown by the solid lines is simply a result of a fixed initial gap position. Finally, while we showed that the large spread in the central radius of \textit{exoKuiper} belts could be obtained by starting the gap at different locations, reproducing the radius distribution is, however, beyond the scope of this paper. In addition to the large spread in radii, the radius distribution of exoKuiper belts is known to be a function of stellar luminosity \citep{Matra2018mmlaw}, with belts being on average larger around more luminous and massive stars. Therefore, not only the large spread in radii needs to be explained, but also its dependence on its host star. This dependency could be related to how protoplanetary disks are as well on average larger and contain more dust around more massive stars \citep{Andrews2013, Pascucci2016, Andrews2018sizes}, and to where dust traps are more likely to form.

\subsection{Outward migrating dust trap}
\begin{figure}
    \centering
    \includegraphics[width=\columnwidth]{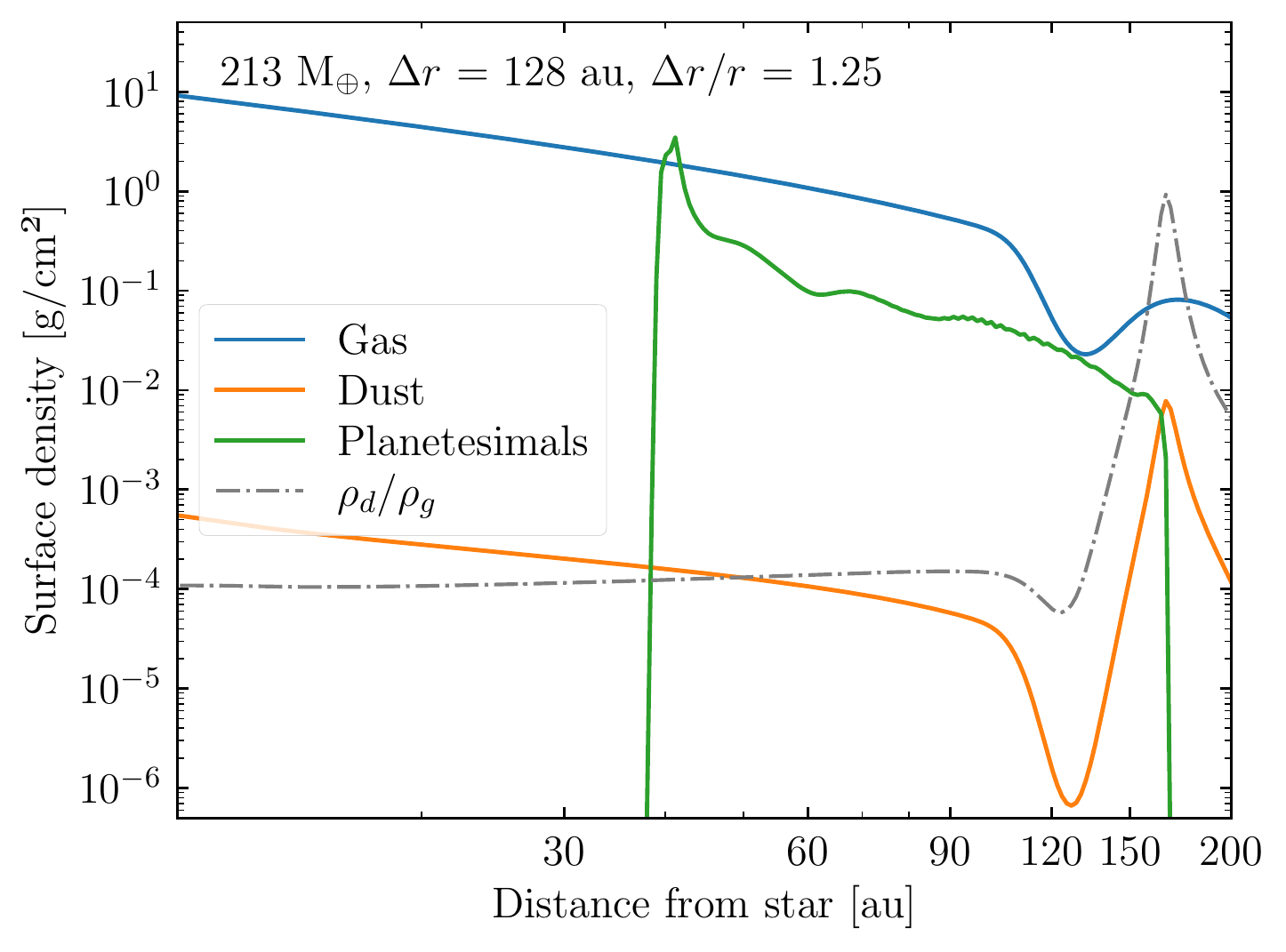}
    \caption{Outward migrating prime case gap ($\alpha = 10^{-3}$, $A = 10$, $f = 100\%$) initially at 30 au evolved for 10 Myr. The gap migrates 100 au in this time and produces the widest belt of all simulation results.}
    \label{fig:outward}
\end{figure}

Finally, we explored a dust trap migrating outward. While inward migration is what is mostly expected if the gap was caused by a planet undergoing type II migration, there are special cases where the dust trap could be moving outward. One of such cases is the scenario in which two massive planets are locked in the 3:2 resonance, leading to an outward migration mantaining the 3:2 resonance \citep{Masset2001, Crida2009}. Under especial circumstances, single massive planets can also migrate outwards \citep[e.g.][]{Masset2003, Crida2007, Peplinski2008,  Lin2012, Dempsey2021}, although these cases might be inconsistent with our assumption of an axisymmetric disc. It could also be the case that the dust traps are caused by mechanisms that do not involve the presence of planets, and that these could migrate as discs evolve. 
For example, if the dust trap is linked to non-ideal MHD effects \citep{Flock2015}, its location could move outwards due to the viscous spreading on a viscous timescale. The dead-zone outer edge is particularly sensitive to the radial profile of the gas surface density \citep[][Delage et al. 2021 submitted]{Dzyurkevich2013}, and thus this location could move outwards as the disc viscously spreads. Therefore, for completeness we decide to explore the outward migrating case even though it might be unlikely.  We simulated the prime case gap starting at 30 au  and travelling at the nominal (opposite) velocity for 10 Myr. Figure~\ref{fig:outward} shows the gas, dust and planetesimal surface densities along with the resulting belt characteristics. Note that the very small crinkles appearing at > 80 au are a numerical artefact.




The planetesimal surface density magnitude decays over time, due to the decreasing availability of dust at higher disc radii and the past planetesimal formation. The resulting belt is actually the widest of all our results, simply because initially planetesimals form rapidly at the closer central position of 30 au, and unlike in the other prime case simulations, the gap does not run into the inner simulated disc radius of 10 au but can migrate outward for the full simulation time of 10~Myr.

An outward migrating gap might not necessarily lead to a wide planetesimal belt if the gap is produced by a planet. An outward migrating planet would sweep the planetesimals that have formed exterior to its orbit, either accreting or scattering them to larger or smaller radii. Depending on the migration speed, planetesimal surface density and presence of inner planets, planetesimal scattering could enhance or slow down its migration \citep{Ida2000, Kirsh2009, Morrison_2018}. However, planetesimal scattering could as well widen the planetesimal belt as scattered planetesimals are implanted onto larger or smaller radii \citep[e.g.][]{Walsh2011}.

\section{Conclusions}
\label{sec:conclusion}
In this work we have explored how migrating dust traps in protoplanetary discs can lead to the creation of wide planetesimal belts, serving as a first step to link \textit{exoKuiper} belts and rings in protoplanetary discs. Using the state-of-the-art dust evolution software \textsc{DustPy}, we investigated the initial conditions most favourable to planetesimal formation for a stationary dust trap, and extended this model to a migrating trap. The main findings of this paper are summarised below.
\begin{enumerate}
    \item Planetesimal formation is most favourable in low viscosity ($\alpha = 10^{-4}$) discs with steep dust traps close to the central star to maximise dust trapping and trigger the streaming instability. We find these conditions are superior at forming planetesimals in both efficiency and timescale.
    \item If the initial disc and trap conditions are favourable to planetesimal formation, dust can still effectively accumulate and form planetesimals as the trap migrates. This leads to a positive correlation between the inward radial speed and resulting planetesimal belt width. 
    \item If the initial disc and gap conditions are not favourable for a stationary dust trap, we find that moving the trap did not improve these conditions and planetesimal formation remained inefficient.
    \item The large widths of observed belts constrain $\alpha$ to values $\geq 4\times10^{-4}$ at tens of au, if 50~au wide planetesimal belts are indeed formed by migrating dust traps over 10~Myr and not external effects reducing the gas surface density (e.g. photoevaporation). Traps in $\alpha = 10^{-4}$ discs move too slowly to form wide planetesimal belts due to the proportionality between $\alpha$ and inward radial velocity. Wider discs or shorter migration timescales would require even higher values of $\alpha$.
    \item The large spread in the widths and radii of exoKuiper belts could be due to different trap migration speeds (or protoplanetary disc lifetimes) and different starting locations, respectively.
\end{enumerate}

\section*{Acknowledgements}
We would like to thank Yuhito Shibaike for a very thorough and constructive review of the paper that improved its quality and clarity. We would also like to thank Mario Flock for providing valuable comments. The work presented in this paper began as a summer student project at the Max Planck Institute for Astronomy, and EM and SM would like to thank the institute for facilitating the internship program. Thanks is also extended to the MPIA IT department for continued access and support with the CPU cluster. SM is supported by a Junior Research Fellowship from Jesus College, University of Cambridge. PP acknowledges support provided by the Alexander von Humboldt Foundation in the framework of the Sofja Kovalevskaja Award endowed by the Federal Ministry of Education and Research. SS and TB acknowledge funding from the European Research Council (ERC) under the European Union's Horizon 2020 research and innovation programme under grant agreement No 714769 and funding by the Deutsche Forschungsgemeinschaft (DFG, German Research Foundation) under Germany's Excellence Strategy - EXC-2094 - 390783311 and under Ref no. FOR 2634/1.

\section*{Data Availability}

The data underlying this article will be shared on reasonable request to the corresponding author. \textsc{DustPy} is available at https://stammler.github.io/dustpy/.



\bibliographystyle{mnras}
\bibliography{biblio} 

\begin{thebibliography}{}
\makeatletter
\relax
\def\mn@urlcharsother{\let\do\@makeother \do\$\do\&\do\#\do\^\do\_\do\%\do\~}
\def\mn@doi{\begingroup\mn@urlcharsother \@ifnextchar [ {\mn@doi@}
  {\mn@doi@[]}}
\def\mn@doi@[#1]#2{\def\@tempa{#1}\ifx\@tempa\@empty \href
  {http://dx.doi.org/#2} {doi:#2}\else \href {http://dx.doi.org/#2} {#1}\fi
  \endgroup}
\def\mn@eprint#1#2{\mn@eprint@#1:#2::\@nil}
\def\mn@eprint@arXiv#1{\href {http://arxiv.org/abs/#1} {{\tt arXiv:#1}}}
\def\mn@eprint@dblp#1{\href {http://dblp.uni-trier.de/rec/bibtex/#1.xml}
  {dblp:#1}}
\def\mn@eprint@#1:#2:#3:#4\@nil{\def\@tempa {#1}\def\@tempb {#2}\def\@tempc
  {#3}\ifx \@tempc \@empty \let \@tempc \@tempb \let \@tempb \@tempa \fi \ifx
  \@tempb \@empty \def\@tempb {arXiv}\fi \@ifundefined
  {mn@eprint@\@tempb}{\@tempb:\@tempc}{\expandafter \expandafter \csname
  mn@eprint@\@tempb\endcsname \expandafter{\@tempc}}}

\bibitem[\protect\citeauthoryear{{Andrews}, {Rosenfeld}, {Kraus}  \&
  {Wilner}}{{Andrews} et~al.}{2013}]{Andrews2013}
{Andrews} S.~M.,  {Rosenfeld} K.~A.,  {Kraus} A.~L.,   {Wilner} D.~J.,  2013,
  \mn@doi [\apj] {10.1088/0004-637X/771/2/129}, \href
  {https://ui.adsabs.harvard.edu/abs/2013ApJ...771..129A} {771, 129}

\bibitem[\protect\citeauthoryear{{Andrews}, {Terrell}, {Tripathi}, {Ansdell},
  {Williams}  \& {Wilner}}{{Andrews} et~al.}{2018a}]{Andrews2018sizes}
{Andrews} S.~M.,  {Terrell} M.,  {Tripathi} A.,  {Ansdell} M.,  {Williams}
  J.~P.,   {Wilner} D.~J.,  2018a, \mn@doi [\apj] {10.3847/1538-4357/aadd9f},
  \href {https://ui.adsabs.harvard.edu/abs/2018ApJ...865..157A} {865, 157}

\bibitem[\protect\citeauthoryear{{Andrews} et~al.,}{{Andrews}
  et~al.}{2018b}]{Andrews2018}
{Andrews} S.~M.,  et~al., 2018b, \mn@doi [\apjl] {10.3847/2041-8213/aaf741},
  \href {https://ui.adsabs.harvard.edu/abs/2018ApJ...869L..41A} {869, L41}

\bibitem[\protect\citeauthoryear{{Baruteau} et~al.,}{{Baruteau}
  et~al.}{2014}]{Baruteau2014}
{Baruteau} C.,  et~al., 2014, in {Beuther} H.,  {Klessen} R.~S.,  {Dullemond}
  C.~P.,   {Henning} T.,  eds, Protostars and Planets VI. p.~667 (\mn@eprint
  {arXiv} {1312.4293}), \mn@doi{10.2458/azu\_uapress\_9780816531240-ch029}

\bibitem[\protect\citeauthoryear{{Birnstiel}, {Dullemond}  \&
  {Brauer}}{{Birnstiel} et~al.}{2010}]{Birnstiel_2010}
{Birnstiel} T.,  {Dullemond} C.~P.,   {Brauer} F.,  2010, \mn@doi [\aap]
  {10.1051/0004-6361/200913731}, \href
  {https://ui.adsabs.harvard.edu/abs/2010A&A...513A..79B} {513, A79}

\bibitem[\protect\citeauthoryear{{Blum} \& {Wurm}}{{Blum} \&
  {Wurm}}{2000}]{Blum2000}
{Blum} J.,  {Wurm} G.,  2000, \mn@doi [\icarus] {10.1006/icar.1999.6234}, \href
  {https://ui.adsabs.harvard.edu/abs/2000Icar..143..138B} {143, 138}

\bibitem[\protect\citeauthoryear{{Booth} et~al.,}{{Booth}
  et~al.}{2017}]{Booth2017}
{Booth} M.,  et~al., 2017, \mn@doi [\mnras] {10.1093/mnras/stx1072}, \href
  {https://ui.adsabs.harvard.edu/abs/2017MNRAS.469.3200B} {469, 3200}

\bibitem[\protect\citeauthoryear{{Brauer}, {Dullemond}  \& {Henning}}{{Brauer}
  et~al.}{2008}]{Brauer_2008}
{Brauer} F.,  {Dullemond} C.~P.,   {Henning} T.,  2008, \mn@doi [\aap]
  {10.1051/0004-6361:20077759}, \href
  {https://ui.adsabs.harvard.edu/abs/2008A&A...480..859B} {480, 859}

\bibitem[\protect\citeauthoryear{{Carrera}, {Gorti}, {Johansen}  \&
  {Davies}}{{Carrera} et~al.}{2017}]{Carrera2017}
{Carrera} D.,  {Gorti} U.,  {Johansen} A.,   {Davies} M.~B.,  2017, \mn@doi
  [\apj] {10.3847/1538-4357/aa6932}, \href
  {https://ui.adsabs.harvard.edu/abs/2017ApJ...839...16C} {839, 16}

\bibitem[\protect\citeauthoryear{{Carrera}, {Simon}, {Li}, {Kretke}  \&
  {Klahr}}{{Carrera} et~al.}{2021}]{Carrera2021}
{Carrera} D.,  {Simon} J.~B.,  {Li} R.,  {Kretke} K.~A.,   {Klahr} H.,  2021,
  \mn@doi [\aj] {10.3847/1538-3881/abd4d9}, \href
  {https://ui.adsabs.harvard.edu/abs/2021AJ....161...96C} {161, 96}

\bibitem[\protect\citeauthoryear{{Cieza} et~al.,}{{Cieza}
  et~al.}{2021}]{Cieza2021}
{Cieza} L.~A.,  et~al., 2021, \mn@doi [\mnras] {10.1093/mnras/staa3787}, \href
  {https://ui.adsabs.harvard.edu/abs/2021MNRAS.501.2934C} {501, 2934}

\bibitem[\protect\citeauthoryear{{Crida} \& {Morbidelli}}{{Crida} \&
  {Morbidelli}}{2007}]{Crida2007}
{Crida} A.,  {Morbidelli} A.,  2007, \mn@doi [\mnras]
  {10.1111/j.1365-2966.2007.11704.x}, \href
  {https://ui.adsabs.harvard.edu/abs/2007MNRAS.377.1324C} {377, 1324}

\bibitem[\protect\citeauthoryear{{Crida}, {Masset}  \& {Morbidelli}}{{Crida}
  et~al.}{2009}]{Crida2009}
{Crida} A.,  {Masset} F.,   {Morbidelli} A.,  2009, \mn@doi [\apjl]
  {10.1088/0004-637X/705/2/L148}, \href
  {https://ui.adsabs.harvard.edu/abs/2009ApJ...705L.148C} {705, L148}

\bibitem[\protect\citeauthoryear{{Daley} et~al.,}{{Daley}
  et~al.}{2019}]{Daley2019}
{Daley} C.,  et~al., 2019, \mn@doi [\apj] {10.3847/1538-4357/ab1074}, \href
  {https://ui.adsabs.harvard.edu/abs/2019ApJ...875...87D} {875, 87}

\bibitem[\protect\citeauthoryear{{Dempsey}, {Mu{\~n}oz}  \&
  {Lithwick}}{{Dempsey} et~al.}{2021}]{Dempsey2021}
{Dempsey} A.~M.,  {Mu{\~n}oz} D.~J.,   {Lithwick} Y.,  2021, arXiv e-prints,
  \href {https://ui.adsabs.harvard.edu/abs/2021arXiv210505277D} {p.
  arXiv:2105.05277}

\bibitem[\protect\citeauthoryear{{Dipierro}, {Price}, {Laibe}, {Hirsh},
  {Cerioli}  \& {Lodato}}{{Dipierro} et~al.}{2015}]{Dipierro2015}
{Dipierro} G.,  {Price} D.,  {Laibe} G.,  {Hirsh} K.,  {Cerioli} A.,   {Lodato}
  G.,  2015, \mn@doi [\mnras] {10.1093/mnrasl/slv105}, \href
  {http://adsabs.harvard.edu/abs/2015MNRAS.453L..73D} {453, L73}

\bibitem[\protect\citeauthoryear{{Dong}, {Li}, {Chiang}  \& {Li}}{{Dong}
  et~al.}{2017}]{Dong2017}
{Dong} R.,  {Li} S.,  {Chiang} E.,   {Li} H.,  2017, \mn@doi [\apj]
  {10.3847/1538-4357/aa72f2}, \href
  {http://adsabs.harvard.edu/abs/2017ApJ...843..127D} {843, 127}

\bibitem[\protect\citeauthoryear{{Dr{\k{a}}{\.z}kowska}, {Alibert}  \&
  {Moore}}{{Dr{\k{a}}{\.z}kowska} et~al.}{2016}]{Drazkowska2016}
{Dr{\k{a}}{\.z}kowska} J.,  {Alibert} Y.,   {Moore} B.,  2016, \mn@doi [\aap]
  {10.1051/0004-6361/201628983}, \href
  {https://ui.adsabs.harvard.edu/abs/2016A&A...594A.105D} {594, A105}

\bibitem[\protect\citeauthoryear{{Dr{\k{a}}{\.z}kowska}, {Stammler}  \&
  {Birnstiel}}{{Dr{\k{a}}{\.z}kowska} et~al.}{2021}]{Dr_kowska_2021}
{Dr{\k{a}}{\.z}kowska} J.,  {Stammler} S.~M.,   {Birnstiel} T.,  2021, \mn@doi
  [\aap] {10.1051/0004-6361/202039925}, \href
  {https://ui.adsabs.harvard.edu/abs/2021A&A...647A..15D} {647, A15}

\bibitem[\protect\citeauthoryear{{Dullemond} \& {Penzlin}}{{Dullemond} \&
  {Penzlin}}{2018}]{Dullemond2018a}
{Dullemond} C.~P.,  {Penzlin} A.~B.~T.,  2018, \mn@doi [\aap]
  {10.1051/0004-6361/201731878}, \href
  {http://adsabs.harvard.edu/abs/2018A%26A...609A..50D} {609, A50}

\bibitem[\protect\citeauthoryear{{Dullemond} et~al.,}{{Dullemond}
  et~al.}{2018}]{Dullemond_2018}
{Dullemond} C.~P.,  et~al., 2018, \mn@doi [\apjl] {10.3847/2041-8213/aaf742},
  \href {https://ui.adsabs.harvard.edu/abs/2018ApJ...869L..46D} {869, L46}

\bibitem[\protect\citeauthoryear{{Dzyurkevich}, {Turner}, {Henning}  \&
  {Kley}}{{Dzyurkevich} et~al.}{2013}]{Dzyurkevich2013}
{Dzyurkevich} N.,  {Turner} N.~J.,  {Henning} T.,   {Kley} W.,  2013, \mn@doi
  [\apj] {10.1088/0004-637X/765/2/114}, \href
  {https://ui.adsabs.harvard.edu/abs/2013ApJ...765..114D} {765, 114}

\bibitem[\protect\citeauthoryear{{Ercolano}, {Jennings}, {Rosotti}  \&
  {Birnstiel}}{{Ercolano} et~al.}{2017}]{Ercolano2017}
{Ercolano} B.,  {Jennings} J.,  {Rosotti} G.,   {Birnstiel} T.,  2017, \mn@doi
  [\mnras] {10.1093/mnras/stx2294}, \href
  {https://ui.adsabs.harvard.edu/abs/2017MNRAS.472.4117E} {472, 4117}

\bibitem[\protect\citeauthoryear{{Eriksson}, {Johansen}  \& {Liu}}{{Eriksson}
  et~al.}{2020}]{Eriksson2020}
{Eriksson} L. E.~J.,  {Johansen} A.,   {Liu} B.,  2020, \mn@doi [\aap]
  {10.1051/0004-6361/201937037}, \href
  {https://ui.adsabs.harvard.edu/abs/2020A&A...635A.110E} {635, A110}

\bibitem[\protect\citeauthoryear{{Eriksson}, {Ronnet}  \&
  {Johansen}}{{Eriksson} et~al.}{2021}]{Eriksson2021}
{Eriksson} L. E.~J.,  {Ronnet} T.,   {Johansen} A.,  2021, \mn@doi [\aap]
  {10.1051/0004-6361/202039889}, \href
  {https://ui.adsabs.harvard.edu/abs/2021A&A...648A.112E} {648, A112}

\bibitem[\protect\citeauthoryear{{Faramaz} et~al.,}{{Faramaz}
  et~al.}{2019}]{Faramaz2019}
{Faramaz} V.,  et~al., 2019, \mn@doi [\aj] {10.3847/1538-3881/ab3ec1}, \href
  {https://ui.adsabs.harvard.edu/abs/2019AJ....158..162F} {158, 162}

\bibitem[\protect\citeauthoryear{{Fedele}, {van den Ancker}, {Henning},
  {Jayawardhana}  \& {Oliveira}}{{Fedele} et~al.}{2010}]{fedele_2010}
{Fedele} D.,  {van den Ancker} M.~E.,  {Henning} T.,  {Jayawardhana} R.,
  {Oliveira} J.~M.,  2010, \mn@doi [\aap] {10.1051/0004-6361/200912810}, \href
  {https://ui.adsabs.harvard.edu/abs/2010A&A...510A..72F} {510, A72}

\bibitem[\protect\citeauthoryear{{Fernandez} \& {Ip}}{{Fernandez} \&
  {Ip}}{1984}]{Fernandez1984}
{Fernandez} J.~A.,  {Ip} W.-H.,  1984, \mn@doi [\icarus]
  {10.1016/0019-1035(84)90101-5}, \href
  {http://adsabs.harvard.edu/abs/1984Icar...58..109F} {58, 109}

\bibitem[\protect\citeauthoryear{{Flaherty}, {Hughes}, {Rosenfeld}, {Andrews},
  {Chiang}, {Simon}, {Kerzner}  \& {Wilner}}{{Flaherty}
  et~al.}{2015}]{Flaherty2015}
{Flaherty} K.~M.,  {Hughes} A.~M.,  {Rosenfeld} K.~A.,  {Andrews} S.~M.,
  {Chiang} E.,  {Simon} J.~B.,  {Kerzner} S.,   {Wilner} D.~J.,  2015, \mn@doi
  [\apj] {10.1088/0004-637X/813/2/99}, \href
  {https://ui.adsabs.harvard.edu/abs/2015ApJ...813...99F} {813, 99}

\bibitem[\protect\citeauthoryear{{Flaherty} et~al.,}{{Flaherty}
  et~al.}{2017}]{Flaherty2017}
{Flaherty} K.~M.,  et~al., 2017, \mn@doi [\apj] {10.3847/1538-4357/aa79f9},
  \href {https://ui.adsabs.harvard.edu/abs/2017ApJ...843..150F} {843, 150}

\bibitem[\protect\citeauthoryear{{Flaherty} et~al.,}{{Flaherty}
  et~al.}{2020}]{Flaherty2020}
{Flaherty} K.,  et~al., 2020, \mn@doi [\apj] {10.3847/1538-4357/ab8cc5}, \href
  {https://ui.adsabs.harvard.edu/abs/2020ApJ...895..109F} {895, 109}

\bibitem[\protect\citeauthoryear{{Flock}, {Ruge}, {Dzyurkevich}, {Henning},
  {Klahr}  \& {Wolf}}{{Flock} et~al.}{2015}]{Flock2015}
{Flock} M.,  {Ruge} J.~P.,  {Dzyurkevich} N.,  {Henning} T.,  {Klahr} H.,
  {Wolf} S.,  2015, \mn@doi [\aap] {10.1051/0004-6361/201424693}, \href
  {https://ui.adsabs.harvard.edu/#abs/2015A&A...574A..68F} {574, A68}

\bibitem[\protect\citeauthoryear{{G{\'a}rate}, {Birnstiel},
  {Dr{\k{a}}{\.z}kowska}  \& {Stammler}}{{G{\'a}rate} et~al.}{2020}]{garate}
{G{\'a}rate} M.,  {Birnstiel} T.,  {Dr{\k{a}}{\.z}kowska} J.,   {Stammler}
  S.~M.,  2020, \mn@doi [\aap] {10.1051/0004-6361/201936067}, \href
  {https://ui.adsabs.harvard.edu/abs/2020A&A...635A.149G} {635, A149}

\bibitem[\protect\citeauthoryear{{Gomes}, {Morbidelli}  \& {Levison}}{{Gomes}
  et~al.}{2004}]{Gomes2004}
{Gomes} R.~S.,  {Morbidelli} A.,   {Levison} H.~F.,  2004, \mn@doi [\icarus]
  {10.1016/j.icarus.2004.03.011}, \href
  {https://ui.adsabs.harvard.edu/abs/2004Icar..170..492G} {170, 492}

\bibitem[\protect\citeauthoryear{{Gundlach} \& {Blum}}{{Gundlach} \&
  {Blum}}{2015}]{Gundlach2015}
{Gundlach} B.,  {Blum} J.,  2015, \mn@doi [\apj] {10.1088/0004-637X/798/1/34},
  \href {https://ui.adsabs.harvard.edu/abs/2015ApJ...798...34G} {798, 34}

\bibitem[\protect\citeauthoryear{{Gundlach} et~al.,}{{Gundlach}
  et~al.}{2018}]{gundlach_2018}
{Gundlach} B.,  et~al., 2018, \mn@doi [\mnras] {10.1093/mnras/sty1550}, \href
  {https://ui.adsabs.harvard.edu/abs/2018MNRAS.479.1273G} {479, 1273}

\bibitem[\protect\citeauthoryear{{Haisch}, {Lada}  \& {Lada}}{{Haisch}
  et~al.}{2001}]{Haisch_Jr__2001}
{Haisch} Karl~E. J.,  {Lada} E.~A.,   {Lada} C.~J.,  2001, \mn@doi [\apjl]
  {10.1086/320685}, \href
  {https://ui.adsabs.harvard.edu/abs/2001ApJ...553L.153H} {553, L153}

\bibitem[\protect\citeauthoryear{{Ida}, {Bryden}, {Lin}  \& {Tanaka}}{{Ida}
  et~al.}{2000}]{Ida2000}
{Ida} S.,  {Bryden} G.,  {Lin} D.~N.~C.,   {Tanaka} H.,  2000, \mn@doi [\apj]
  {10.1086/308720}, \href
  {https://ui.adsabs.harvard.edu/abs/2000ApJ...534..428I} {534, 428}

\bibitem[\protect\citeauthoryear{{Jiang} \& {Ormel}}{{Jiang} \&
  {Ormel}}{2021}]{Jiang2021}
{Jiang} H.,  {Ormel} C.~W.,  2021, \mn@doi [\mnras] {10.1093/mnras/stab1278},
  \href {https://ui.adsabs.harvard.edu/abs/2021MNRAS.505.1162J} {505, 1162}

\bibitem[\protect\citeauthoryear{{Johansen}, {Oishi}, {Mac Low}, {Klahr},
  {Henning}  \& {Youdin}}{{Johansen} et~al.}{2007}]{Johansen2007}
{Johansen} A.,  {Oishi} J.~S.,  {Mac Low} M.-M.,  {Klahr} H.,  {Henning} T.,
  {Youdin} A.,  2007, \mn@doi [\nat] {10.1038/nature06086}, \href
  {http://adsabs.harvard.edu/abs/2007Natur.448.1022J} {448, 1022}

\bibitem[\protect\citeauthoryear{{Johansen}, {Youdin}  \& {Klahr}}{{Johansen}
  et~al.}{2009}]{Johansen_2009}
{Johansen} A.,  {Youdin} A.,   {Klahr} H.,  2009, \mn@doi [\apj]
  {10.1088/0004-637X/697/2/1269}, \href
  {https://ui.adsabs.harvard.edu/abs/2009ApJ...697.1269J} {697, 1269}

\bibitem[\protect\citeauthoryear{{Kennedy}, {Marino}, {Matr{\`a}}, {Pani{\'c}},
  {Wilner}, {Wyatt}  \& {Yelverton}}{{Kennedy} et~al.}{2018}]{Kennedy2018}
{Kennedy} G.~M.,  {Marino} S.,  {Matr{\`a}} L.,  {Pani{\'c}} O.,  {Wilner} D.,
  {Wyatt} M.~C.,   {Yelverton} B.,  2018, \mn@doi [\mnras]
  {10.1093/mnras/sty135}, \href
  {http://adsabs.harvard.edu/abs/2018MNRAS.475.4924K} {475, 4924}

\bibitem[\protect\citeauthoryear{{Kirsh}, {Duncan}, {Brasser}  \&
  {Levison}}{{Kirsh} et~al.}{2009}]{Kirsh2009}
{Kirsh} D.~R.,  {Duncan} M.,  {Brasser} R.,   {Levison} H.~F.,  2009, \mn@doi
  [\icarus] {10.1016/j.icarus.2008.05.028}, \href
  {https://ui.adsabs.harvard.edu/abs/2009Icar..199..197K} {199, 197}

\bibitem[\protect\citeauthoryear{{Klahr} \& {Schreiber}}{{Klahr} \&
  {Schreiber}}{2020}]{Klahr_2020}
{Klahr} H.,  {Schreiber} A.,  2020, \mn@doi [\apj] {10.3847/1538-4357/abac58},
  \href {https://ui.adsabs.harvard.edu/abs/2020ApJ...901...54K} {901, 54}

\bibitem[\protect\citeauthoryear{{Kley} \& {Nelson}}{{Kley} \&
  {Nelson}}{2012}]{Kley2012}
{Kley} W.,  {Nelson} R.~P.,  2012, \mn@doi [\araa]
  {10.1146/annurev-astro-081811-125523}, \href
  {https://ui.adsabs.harvard.edu/abs/2012ARA&A..50..211K} {50, 211}

\bibitem[\protect\citeauthoryear{{Lenz}, {Klahr}  \& {Birnstiel}}{{Lenz}
  et~al.}{2019}]{Lenz2019}
{Lenz} C.~T.,  {Klahr} H.,   {Birnstiel} T.,  2019, \mn@doi [\apj]
  {10.3847/1538-4357/ab05d9}, \href
  {https://ui.adsabs.harvard.edu/abs/2019ApJ...874...36L} {874, 36}

\bibitem[\protect\citeauthoryear{{Lenz}, {Klahr}, {Birnstiel}, {Kretke}  \&
  {Stammler}}{{Lenz} et~al.}{2020}]{Lenz_2020}
{Lenz} C.~T.,  {Klahr} H.,  {Birnstiel} T.,  {Kretke} K.,   {Stammler} S.,
  2020, \mn@doi [\aap] {10.1051/0004-6361/202037878}, \href
  {https://ui.adsabs.harvard.edu/abs/2020A&A...640A..61L} {640, A61}

\bibitem[\protect\citeauthoryear{{Li} \& {Youdin}}{{Li} \&
  {Youdin}}{2021}]{Li2021}
{Li} R.,  {Youdin} A.,  2021, arXiv e-prints, \href
  {https://ui.adsabs.harvard.edu/abs/2021arXiv210506042L} {p. arXiv:2105.06042}

\bibitem[\protect\citeauthoryear{{Li}, {Lubow}, {Li}  \& {Lin}}{{Li}
  et~al.}{2009}]{Li2009}
{Li} H.,  {Lubow} S.~H.,  {Li} S.,   {Lin} D.~N.~C.,  2009, \mn@doi [\apjl]
  {10.1088/0004-637X/690/1/L52}, \href
  {https://ui.adsabs.harvard.edu/abs/2009ApJ...690L..52L} {690, L52}

\bibitem[\protect\citeauthoryear{{Lin} \& {Papaloizou}}{{Lin} \&
  {Papaloizou}}{1979}]{Lin1979}
{Lin} D.~N.~C.,  {Papaloizou} J.,  1979, \mn@doi [\mnras]
  {10.1093/mnras/186.4.799}, \href
  {https://ui.adsabs.harvard.edu/abs/1979MNRAS.186..799L} {186, 799}

\bibitem[\protect\citeauthoryear{{Lin} \& {Papaloizou}}{{Lin} \&
  {Papaloizou}}{1986}]{Lin1986}
{Lin} D.~N.~C.,  {Papaloizou} J.,  1986, \mn@doi [\apj] {10.1086/164653}, \href
  {https://ui.adsabs.harvard.edu/abs/1986ApJ...309..846L} {309, 846}

\bibitem[\protect\citeauthoryear{{Lin} \& {Papaloizou}}{{Lin} \&
  {Papaloizou}}{2012}]{Lin2012}
{Lin} M.-K.,  {Papaloizou} J. C.~B.,  2012, \mn@doi [\mnras]
  {10.1111/j.1365-2966.2011.20352.x}, \href
  {https://ui.adsabs.harvard.edu/abs/2012MNRAS.421..780L} {421, 780}

\bibitem[\protect\citeauthoryear{{Long} et~al.,}{{Long}
  et~al.}{2018}]{Long2018}
{Long} F.,  et~al., 2018, \mn@doi [\apj] {10.3847/1538-4357/aae8e1}, \href
  {https://ui.adsabs.harvard.edu/abs/2018ApJ...869...17L} {869, 17}

\bibitem[\protect\citeauthoryear{{Lor{\'e}n-Aguilar} \&
  {Bate}}{{Lor{\'e}n-Aguilar} \& {Bate}}{2015}]{Loren-Aguilar2015}
{Lor{\'e}n-Aguilar} P.,  {Bate} M.~R.,  2015, \mn@doi [\mnras]
  {10.1093/mnrasl/slv109}, \href
  {https://ui.adsabs.harvard.edu/abs/2015MNRAS.453L..78L} {453, L78}

\bibitem[\protect\citeauthoryear{{Lynden-Bell} \& {Pringle}}{{Lynden-Bell} \&
  {Pringle}}{1974}]{LyndenBell1974TheEO}
{Lynden-Bell} D.,  {Pringle} J.~E.,  1974, \mn@doi [\mnras]
  {10.1093/mnras/168.3.603}, \href
  {https://ui.adsabs.harvard.edu/abs/1974MNRAS.168..603L} {168, 603}

\bibitem[\protect\citeauthoryear{{MacGregor} et~al.,}{{MacGregor}
  et~al.}{2017}]{MacGregor2017}
{MacGregor} M.~A.,  et~al., 2017, \mn@doi [\apj] {10.3847/1538-4357/aa71ae},
  \href {http://adsabs.harvard.edu/abs/2017ApJ...842....8M} {842, 8}

\bibitem[\protect\citeauthoryear{{MacGregor} et~al.,}{{MacGregor}
  et~al.}{2019}]{MacGregor2019}
{MacGregor} M.~A.,  et~al., 2019, \mn@doi [\apjl] {10.3847/2041-8213/ab21c2},
  \href {https://ui.adsabs.harvard.edu/abs/2019ApJ...877L..32M} {877, L32}

\bibitem[\protect\citeauthoryear{{Marino} et~al.,}{{Marino}
  et~al.}{2018}]{Marino2018}
{Marino} S.,  et~al., 2018, \mn@doi [\mnras] {10.1093/mnras/sty1790}, \href
  {http://adsabs.harvard.edu/abs/2018MNRAS.479.5423M} {479, 5423}

\bibitem[\protect\citeauthoryear{{Marino}, {Yelverton}, {Booth}, {Faramaz},
  {Kennedy}, {Matr{\`a}}  \& {Wyatt}}{{Marino} et~al.}{2019}]{Marino2019}
{Marino} S.,  {Yelverton} B.,  {Booth} M.,  {Faramaz} V.,  {Kennedy} G.~M.,
  {Matr{\`a}} L.,   {Wyatt} M.~C.,  2019, \mn@doi [\mnras]
  {10.1093/mnras/stz049}, \href
  {https://ui.adsabs.harvard.edu/abs/2019MNRAS.484.1257M} {484, 1257}

\bibitem[\protect\citeauthoryear{{Marino} et~al.,}{{Marino}
  et~al.}{2020}]{Marino2020}
{Marino} S.,  et~al., 2020, \mn@doi [\mnras] {10.1093/mnras/staa2386}, \href
  {https://ui.adsabs.harvard.edu/abs/2020MNRAS.498.1319M} {498, 1319}

\bibitem[\protect\citeauthoryear{{Masset} \& {Papaloizou}}{{Masset} \&
  {Papaloizou}}{2003}]{Masset2003}
{Masset} F.~S.,  {Papaloizou} J.~C.~B.,  2003, \mn@doi [\apj] {10.1086/373892},
  \href {https://ui.adsabs.harvard.edu/abs/2003ApJ...588..494M} {588, 494}

\bibitem[\protect\citeauthoryear{{Masset} \& {Snellgrove}}{{Masset} \&
  {Snellgrove}}{2001}]{Masset2001}
{Masset} F.,  {Snellgrove} M.,  2001, \mn@doi [\mnras]
  {10.1046/j.1365-8711.2001.04159.x}, \href
  {https://ui.adsabs.harvard.edu/abs/2001MNRAS.320L..55M} {320, L55}

\bibitem[\protect\citeauthoryear{{Matr{\`a}}, {Marino}, {Kennedy}, {Wyatt},
  {{\"O}berg}  \& {Wilner}}{{Matr{\`a}} et~al.}{2018}]{Matra2018mmlaw}
{Matr{\`a}} L.,  {Marino} S.,  {Kennedy} G.~M.,  {Wyatt} M.~C.,  {{\"O}berg}
  K.~I.,   {Wilner} D.~J.,  2018, \mn@doi [\apj] {10.3847/1538-4357/aabcc4},
  \href {http://adsabs.harvard.edu/abs/2018ApJ...859...72M} {859, 72}

\bibitem[\protect\citeauthoryear{{Meru}, {Rosotti}, {Booth}, {Nazari}  \&
  {Clarke}}{{Meru} et~al.}{2019}]{Meru2019}
{Meru} F.,  {Rosotti} G.~P.,  {Booth} R.~A.,  {Nazari} P.,   {Clarke} C.~J.,
  2019, \mn@doi [\mnras] {10.1093/mnras/sty2847}, \href
  {https://ui.adsabs.harvard.edu/abs/2019MNRAS.482.3678M} {482, 3678}

\bibitem[\protect\citeauthoryear{{Michel}, {van der Marel}  \&
  {Matthews}}{{Michel} et~al.}{2021}]{Michel2021}
{Michel} A.,  {van der Marel} N.,   {Matthews} B.,  2021, arXiv e-prints, \href
  {https://ui.adsabs.harvard.edu/abs/2021arXiv210405894M} {p. arXiv:2104.05894}

\bibitem[\protect\citeauthoryear{{Morrison} \& {Kratter}}{{Morrison} \&
  {Kratter}}{2018}]{Morrison_2018}
{Morrison} S.~J.,  {Kratter} K.~M.,  2018, \mn@doi [\mnras]
  {10.1093/mnras/sty2657}, \href
  {https://ui.adsabs.harvard.edu/abs/2018MNRAS.481.5180M} {481, 5180}

\bibitem[\protect\citeauthoryear{{Musiolik} \& {Wurm}}{{Musiolik} \&
  {Wurm}}{2019}]{Musiolik_2019}
{Musiolik} G.,  {Wurm} G.,  2019, \mn@doi [\apj] {10.3847/1538-4357/ab0428},
  \href {https://ui.adsabs.harvard.edu/abs/2019ApJ...873...58M} {873, 58}

\bibitem[\protect\citeauthoryear{{Musiolik}, {Teiser}, {Jankowski}  \&
  {Wurm}}{{Musiolik} et~al.}{2016}]{Musiolik2016}
{Musiolik} G.,  {Teiser} J.,  {Jankowski} T.,   {Wurm} G.,  2016, \mn@doi
  [\apj] {10.3847/0004-637X/827/1/63}, \href
  {https://ui.adsabs.harvard.edu/abs/2016ApJ...827...63M} {827, 63}

\bibitem[\protect\citeauthoryear{{Nazari}, {Booth}, {Clarke}, {Rosotti},
  {Tazzari}, {Juhasz}  \& {Meru}}{{Nazari} et~al.}{2019}]{Nazari2019}
{Nazari} P.,  {Booth} R.~A.,  {Clarke} C.~J.,  {Rosotti} G.~P.,  {Tazzari} M.,
  {Juhasz} A.,   {Meru} F.,  2019, \mn@doi [\mnras] {10.1093/mnras/stz836},
  \href {https://ui.adsabs.harvard.edu/abs/2019MNRAS.485.5914N} {485, 5914}

\bibitem[\protect\citeauthoryear{{Nederlander} et~al.,}{{Nederlander}
  et~al.}{2021}]{Nederlander2021}
{Nederlander} A.,  et~al., 2021, arXiv e-prints, \href
  {https://ui.adsabs.harvard.edu/abs/2021arXiv210108849N} {p. arXiv:2101.08849}

\bibitem[\protect\citeauthoryear{{Ormel} \& {Cuzzi}}{{Ormel} \&
  {Cuzzi}}{2007}]{OrmelCuzzi}
{Ormel} C.~W.,  {Cuzzi} J.~N.,  2007, \mn@doi [A\&A]
  {10.1051/0004-6361:20066899}, 466, 413

\bibitem[\protect\citeauthoryear{{Pascucci} et~al.,}{{Pascucci}
  et~al.}{2016}]{Pascucci2016}
{Pascucci} I.,  et~al., 2016, \mn@doi [\apj] {10.3847/0004-637X/831/2/125},
  \href {https://ui.adsabs.harvard.edu/abs/2016ApJ...831..125P} {831, 125}

\bibitem[\protect\citeauthoryear{{Pepli{\'n}ski}, {Artymowicz}  \&
  {Mellema}}{{Pepli{\'n}ski} et~al.}{2008}]{Peplinski2008}
{Pepli{\'n}ski} A.,  {Artymowicz} P.,   {Mellema} G.,  2008, \mn@doi [\mnras]
  {10.1111/j.1365-2966.2008.13339.x}, \href
  {https://ui.adsabs.harvard.edu/abs/2008MNRAS.387.1063P} {387, 1063}

\bibitem[\protect\citeauthoryear{{P{\'e}rez}, {Casassus}, {Baruteau}, {Dong},
  {Hales}  \& {Cieza}}{{P{\'e}rez} et~al.}{2019}]{Perez2019}
{P{\'e}rez} S.,  {Casassus} S.,  {Baruteau} C.,  {Dong} R.,  {Hales} A.,
  {Cieza} L.,  2019, \mn@doi [\aj] {10.3847/1538-3881/ab1f88}, \href
  {https://ui.adsabs.harvard.edu/abs/2019AJ....158...15P} {158, 15}

\bibitem[\protect\citeauthoryear{{Pfalzner}, {Steinhausen}  \&
  {Menten}}{{Pfalzner} et~al.}{2014}]{Pfalzner_2014}
{Pfalzner} S.,  {Steinhausen} M.,   {Menten} K.,  2014, \mn@doi [\apjl]
  {10.1088/2041-8205/793/2/L34}, \href
  {https://ui.adsabs.harvard.edu/abs/2014ApJ...793L..34P} {793, L34}

\bibitem[\protect\citeauthoryear{{Pinilla}, {Benisty}  \&
  {Birnstiel}}{{Pinilla} et~al.}{2012}]{Pinilla2012}
{Pinilla} P.,  {Benisty} M.,   {Birnstiel} T.,  2012, \mn@doi [\aap]
  {10.1051/0004-6361/201219315}, \href
  {http://adsabs.harvard.edu/abs/2012A%26A...545A..81P} {545, A81}

\bibitem[\protect\citeauthoryear{{Pinilla}, {Flock}, {Ovelar}  \&
  {Birnstiel}}{{Pinilla} et~al.}{2016}]{Pinilla2016}
{Pinilla} P.,  {Flock} M.,  {Ovelar} M. d.~J.,   {Birnstiel} T.,  2016, \mn@doi
  [\aap] {10.1051/0004-6361/201628441}, \href
  {https://ui.adsabs.harvard.edu/abs/2016A&A...596A..81P} {596, A81}

\bibitem[\protect\citeauthoryear{{Pinilla}, {Lenz}  \& {Stammler}}{{Pinilla}
  et~al.}{2021}]{Pinilla_2021}
{Pinilla} P.,  {Lenz} C.~T.,   {Stammler} S.~M.,  2021, \mn@doi [\aap]
  {10.1051/0004-6361/202038920}, \href
  {https://ui.adsabs.harvard.edu/abs/2021A&A...645A..70P} {645, A70}

\bibitem[\protect\citeauthoryear{{Pinte}, {Dent}, {M{\'e}nard}, {Hales},
  {Hill}, {Cortes}  \& {de Gregorio-Monsalvo}}{{Pinte}
  et~al.}{2016}]{Pinte2016}
{Pinte} C.,  {Dent} W.~R.~F.,  {M{\'e}nard} F.,  {Hales} A.,  {Hill} T.,
  {Cortes} P.,   {de Gregorio-Monsalvo} I.,  2016, \mn@doi [\apj]
  {10.3847/0004-637X/816/1/25}, \href
  {https://ui.adsabs.harvard.edu/abs/2016ApJ...816...25P} {816, 25}

\bibitem[\protect\citeauthoryear{{Pringle}}{{Pringle}}{1981}]{Pringle_1981}
{Pringle} J.~E.,  1981, \mn@doi [\araa] {10.1146/annurev.aa.19.090181.001033},
  \href {https://ui.adsabs.harvard.edu/abs/1981ARA&A..19..137P} {19, 137}

\bibitem[\protect\citeauthoryear{{Ribas}, {Bouy}  \& {Mer{\'\i}n}}{{Ribas}
  et~al.}{2015}]{ribas_2015}
{Ribas} {\'A}.,  {Bouy} H.,   {Mer{\'\i}n} B.,  2015, \mn@doi [\aap]
  {10.1051/0004-6361/201424846}, \href
  {https://ui.adsabs.harvard.edu/abs/2015A&A...576A..52R} {576, A52}

\bibitem[\protect\citeauthoryear{{Saito} \& {Sirono}}{{Saito} \&
  {Sirono}}{2011}]{Saito2011}
{Saito} E.,  {Sirono} S.-i.,  2011, \mn@doi [\apj]
  {10.1088/0004-637X/728/1/20}, \href
  {https://ui.adsabs.harvard.edu/abs/2011ApJ...728...20S} {728, 20}

\bibitem[\protect\citeauthoryear{{Schoonenberg}, {Ormel}  \&
  {Krijt}}{{Schoonenberg} et~al.}{2018}]{Schoonenberg_2018}
{Schoonenberg} D.,  {Ormel} C.~W.,   {Krijt} S.,  2018, \mn@doi [\aap]
  {10.1051/0004-6361/201834047}, \href
  {https://ui.adsabs.harvard.edu/abs/2018A&A...620A.134S} {620, A134}

\bibitem[\protect\citeauthoryear{{Sellek}, {Booth}  \& {Clarke}}{{Sellek}
  et~al.}{2020}]{Sellek2020}
{Sellek} A.~D.,  {Booth} R.~A.,   {Clarke} C.~J.,  2020, \mn@doi [\mnras]
  {10.1093/mnras/stz3528}, \href
  {https://ui.adsabs.harvard.edu/abs/2020MNRAS.492.1279S} {492, 1279}

\bibitem[\protect\citeauthoryear{{Sepulveda} et~al.,}{{Sepulveda}
  et~al.}{2019}]{Sepulveda2019}
{Sepulveda} A.~G.,  et~al., 2019, \mn@doi [\apj] {10.3847/1538-4357/ab2b98},
  \href {https://ui.adsabs.harvard.edu/abs/2019ApJ...881...84S} {881, 84}

\bibitem[\protect\citeauthoryear{{Shakura} \& {Sunyaev}}{{Shakura} \&
  {Sunyaev}}{1973}]{Shakura_1973}
{Shakura} N.~I.,  {Sunyaev} R.~A.,  1973, \aap, \href
  {https://ui.adsabs.harvard.edu/abs/1973A&A....24..337S} {500, 33}

\bibitem[\protect\citeauthoryear{{Shibaike} \& {Alibert}}{{Shibaike} \&
  {Alibert}}{2020}]{Shibaike2020}
{Shibaike} Y.,  {Alibert} Y.,  2020, \mn@doi [\aap]
  {10.1051/0004-6361/202039086}, \href
  {https://ui.adsabs.harvard.edu/abs/2020A&A...644A..81S} {644, A81}

\bibitem[\protect\citeauthoryear{{Smoluchowski}}{{Smoluchowski}}{1916}]{smol}
{Smoluchowski} M.~V.,  1916, Zeitschrift fur Physik, \href
  {https://ui.adsabs.harvard.edu/abs/1916ZPhy...17..557S} {17, 557}

\bibitem[\protect\citeauthoryear{{Stammler}, {Dr{\k{a}}{\.z}kowska},
  {Birnstiel}, {Klahr}, {Dullemond}  \& {Andrews}}{{Stammler}
  et~al.}{2019}]{Stammler_2019}
{Stammler} S.~M.,  {Dr{\k{a}}{\.z}kowska} J.,  {Birnstiel} T.,  {Klahr} H.,
  {Dullemond} C.~P.,   {Andrews} S.~M.,  2019, \mn@doi [\apjl]
  {10.3847/2041-8213/ab4423}, \href
  {https://ui.adsabs.harvard.edu/abs/2019ApJ...884L...5S} {884, L5}

\bibitem[\protect\citeauthoryear{{Steinpilz}, {Teiser}  \& {Wurm}}{{Steinpilz}
  et~al.}{2019}]{Steinpilz_2019}
{Steinpilz} T.,  {Teiser} J.,   {Wurm} G.,  2019, \mn@doi [\apj]
  {10.3847/1538-4357/ab07bb}, \href
  {https://ui.adsabs.harvard.edu/abs/2019ApJ...874...60S} {874, 60}

\bibitem[\protect\citeauthoryear{{Strubbe} \& {Chiang}}{{Strubbe} \&
  {Chiang}}{2006}]{Strubbe2006}
{Strubbe} L.~E.,  {Chiang} E.~I.,  2006, \mn@doi [\apj] {10.1086/505736}, \href
  {https://ui.adsabs.harvard.edu/abs/2006ApJ...648..652S} {648, 652}

\bibitem[\protect\citeauthoryear{{Takahashi} \& {Inutsuka}}{{Takahashi} \&
  {Inutsuka}}{2014}]{Takahashi2014}
{Takahashi} S.~Z.,  {Inutsuka} S.-i.,  2014, \mn@doi [\apj]
  {10.1088/0004-637X/794/1/55}, \href
  {http://adsabs.harvard.edu/abs/2014ApJ...794...55T} {794, 55}

\bibitem[\protect\citeauthoryear{{Teague} et~al.,}{{Teague}
  et~al.}{2016}]{Teague2016}
{Teague} R.,  et~al., 2016, \mn@doi [\aap] {10.1051/0004-6361/201628550}, \href
  {https://ui.adsabs.harvard.edu/abs/2016A&A...592A..49T} {592, A49}

\bibitem[\protect\citeauthoryear{{Teague}, {Bae}, {Bergin}, {Birnstiel}  \&
  {Foreman-Mackey}}{{Teague} et~al.}{2018a}]{Teague2018kinematics}
{Teague} R.,  {Bae} J.,  {Bergin} E.~A.,  {Birnstiel} T.,   {Foreman-Mackey}
  D.,  2018a, \mn@doi [\apjl] {10.3847/2041-8213/aac6d7}, \href
  {https://ui.adsabs.harvard.edu/abs/2018ApJ...860L..12T} {860, L12}

\bibitem[\protect\citeauthoryear{{Teague} et~al.,}{{Teague}
  et~al.}{2018b}]{Teague2018turbulence}
{Teague} R.,  et~al., 2018b, \mn@doi [\apj] {10.3847/1538-4357/aad80e}, \href
  {https://ui.adsabs.harvard.edu/abs/2018ApJ...864..133T} {864, 133}

\bibitem[\protect\citeauthoryear{{Teague}, {Bae}, {Birnstiel}  \&
  {Bergin}}{{Teague} et~al.}{2018c}]{Teague2018AS209}
{Teague} R.,  {Bae} J.,  {Birnstiel} T.,   {Bergin} E.~A.,  2018c, \mn@doi
  [\apj] {10.3847/1538-4357/aae836}, \href
  {https://ui.adsabs.harvard.edu/abs/2018ApJ...868..113T} {868, 113}

\bibitem[\protect\citeauthoryear{{Teague}, {Bae}  \& {Bergin}}{{Teague}
  et~al.}{2019}]{Teague2019}
{Teague} R.,  {Bae} J.,   {Bergin} E.~A.,  2019, \mn@doi [\nat]
  {10.1038/s41586-019-1642-0}, \href
  {https://ui.adsabs.harvard.edu/abs/2019Natur.574..378T} {574, 378}

\bibitem[\protect\citeauthoryear{{Throop} \& {Bally}}{{Throop} \&
  {Bally}}{2005}]{Throop2005}
{Throop} H.~B.,  {Bally} J.,  2005, \mn@doi [\apjl] {10.1086/430272}, \href
  {https://ui.adsabs.harvard.edu/abs/2005ApJ...623L.149T} {623, L149}

\bibitem[\protect\citeauthoryear{{Uribe}, {Klahr}, {Flock}  \&
  {Henning}}{{Uribe} et~al.}{2011}]{Uribe_2011}
{Uribe} A.~L.,  {Klahr} H.,  {Flock} M.,   {Henning} T.,  2011, \mn@doi [\apj]
  {10.1088/0004-637X/736/2/85}, \href
  {https://ui.adsabs.harvard.edu/abs/2011ApJ...736...85U} {736, 85}

\bibitem[\protect\citeauthoryear{{Villenave} et~al.,}{{Villenave}
  et~al.}{2020}]{Villenave2020}
{Villenave} M.,  et~al., 2020, \mn@doi [\aap] {10.1051/0004-6361/202038087},
  \href {https://ui.adsabs.harvard.edu/abs/2020A&A...642A.164V} {642, A164}

\bibitem[\protect\citeauthoryear{{Wada}, {Tanaka}, {Suyama}, {Kimura}  \&
  {Yamamoto}}{{Wada} et~al.}{2009}]{Wada2009}
{Wada} K.,  {Tanaka} H.,  {Suyama} T.,  {Kimura} H.,   {Yamamoto} T.,  2009,
  \mn@doi [\apj] {10.1088/0004-637X/702/2/1490}, \href
  {https://ui.adsabs.harvard.edu/abs/2009ApJ...702.1490W} {702, 1490}

\bibitem[\protect\citeauthoryear{{Wada}, {Tanaka}, {Suyama}, {Kimura}  \&
  {Yamamoto}}{{Wada} et~al.}{2011}]{Wada2011}
{Wada} K.,  {Tanaka} H.,  {Suyama} T.,  {Kimura} H.,   {Yamamoto} T.,  2011,
  \mn@doi [\apj] {10.1088/0004-637X/737/1/36}, \href
  {https://ui.adsabs.harvard.edu/abs/2011ApJ...737...36W} {737, 36}

\bibitem[\protect\citeauthoryear{{Walsh}, {Morbidelli}, {Raymond}, {O'Brien}
  \& {Mandell}}{{Walsh} et~al.}{2011}]{Walsh2011}
{Walsh} K.~J.,  {Morbidelli} A.,  {Raymond} S.~N.,  {O'Brien} D.~P.,
  {Mandell} A.~M.,  2011, \mn@doi [\nat] {10.1038/nature10201}, \href
  {https://ui.adsabs.harvard.edu/abs/2011Natur.475..206W} {475, 206}

\bibitem[\protect\citeauthoryear{{Whipple}}{{Whipple}}{1972}]{whipple_1972}
{Whipple} F.~L.,  1972, in {Elvius} A.,  ed., From Plasma to Planet. p.~211

\bibitem[\protect\citeauthoryear{{Youdin} \& {Goodman}}{{Youdin} \&
  {Goodman}}{2005}]{Youdin2005}
{Youdin} A.~N.,  {Goodman} J.,  2005, \mn@doi [\apj] {10.1086/426895}, \href
  {http://adsabs.harvard.edu/abs/2005ApJ...620..459Y} {620, 459}

\bibitem[\protect\citeauthoryear{{Youdin} \& {Lithwick}}{{Youdin} \&
  {Lithwick}}{2007}]{Youdin2007ParticleSI}
{Youdin} A.~N.,  {Lithwick} Y.,  2007, \mn@doi [\icarus]
  {10.1016/j.icarus.2007.07.012}, \href
  {https://ui.adsabs.harvard.edu/abs/2007Icar..192..588Y} {192, 588}

\bibitem[\protect\citeauthoryear{{Zhang} et~al.,}{{Zhang}
  et~al.}{2018}]{Zhang2018}
{Zhang} S.,  et~al., 2018, \mn@doi [\apjl] {10.3847/2041-8213/aaf744}, \href
  {https://ui.adsabs.harvard.edu/abs/2018ApJ...869L..47Z} {869, L47}

\makeatother
\end{thebibliography}




\appendix

\section{Surface Density Profile}

\begin{figure}
    \centering
    \includegraphics[width=\columnwidth]{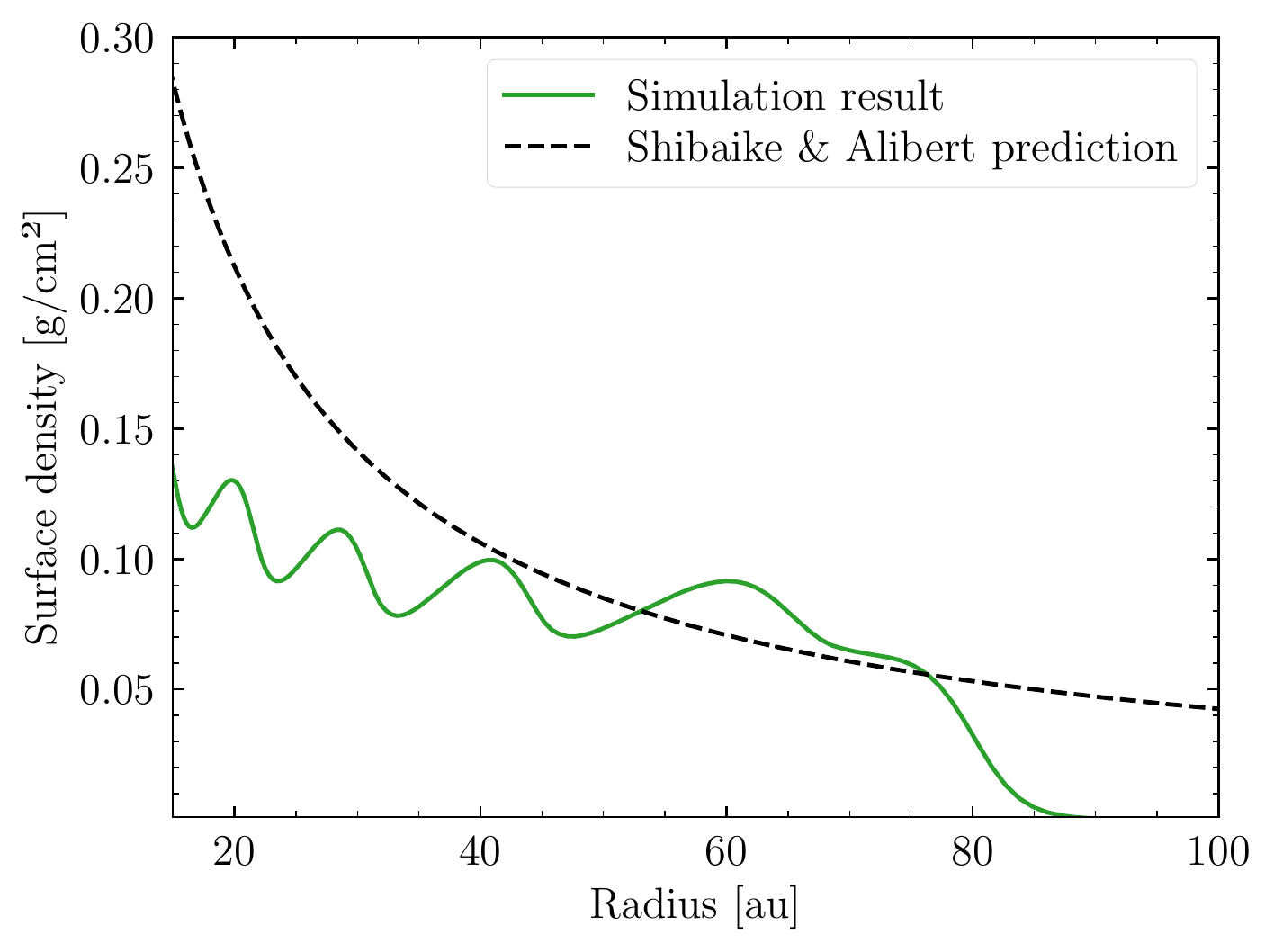}
    \caption{Comparison between simulated planetesimal surface density
      profile of the prime case ($\alpha = 10^{-3}$, $A$ = 10, $f$ =
      100\%), and the theoretical prediction by \citet{Shibaike2020} as
      given in Equation~\ref{eq:shibaike}.}
    \label{fig:slopes}
\end{figure}

\section{Mass Evolution}
\label{appendix_mass}
\begin{figure*}
    \centering
    \includegraphics[width=\linewidth]{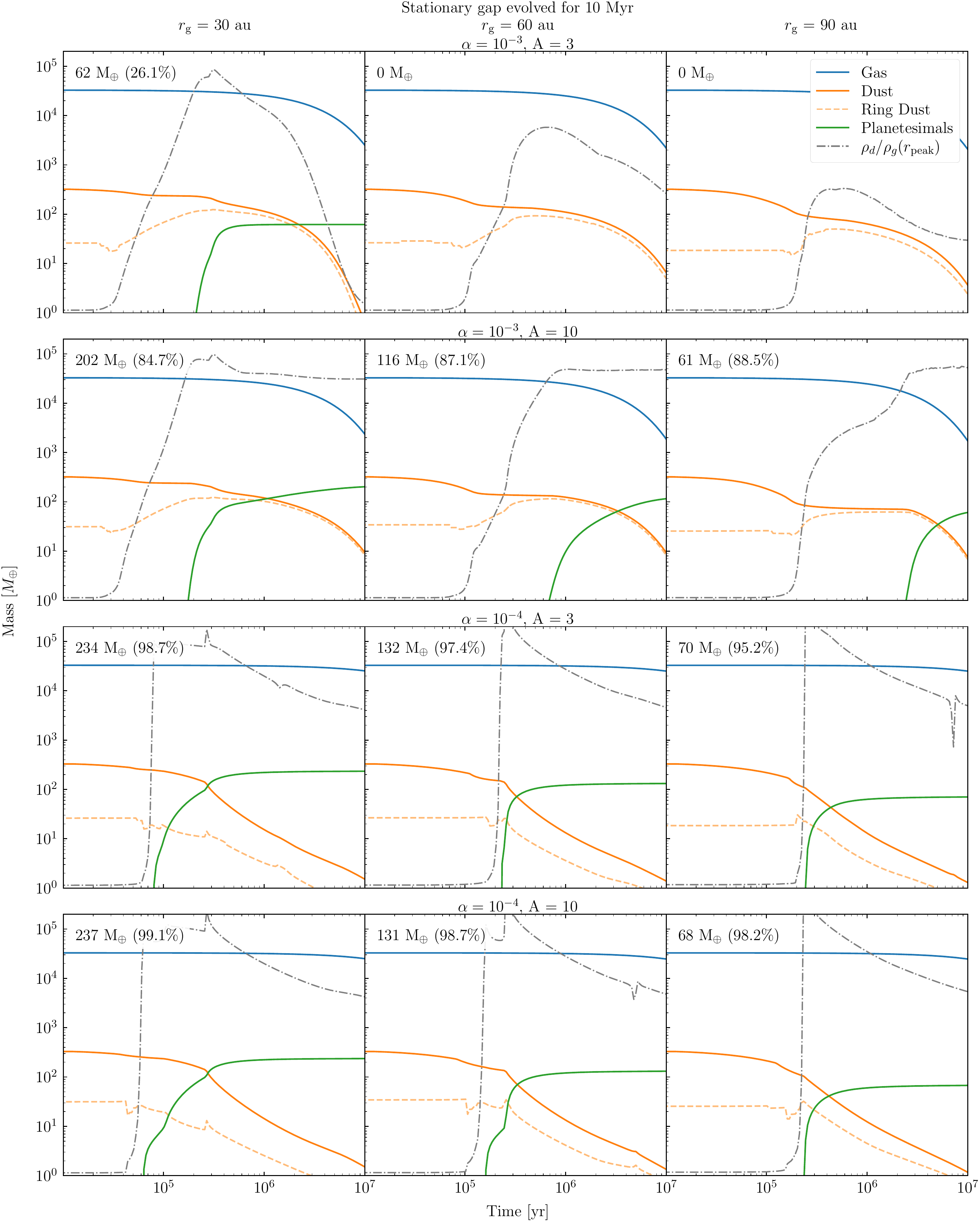}
    \caption{Mass evolution of the gas, dust, ring dust and planetesimals in a protoplanetary disc with a stationary dust trap at gap positions $r_{\rm g}$ = 30, 60 and 90 au with varying viscosity parameter $\alpha$ and gap amplitude $A$. If the gap is at $x = -d$ and dust peak is at $x = 0$, the ring dust was computed by integrating over $x = \pm d/2$. The final planetesimal mass for each simulation is presented, adjacent to the final percentage of initial exterior dust mass transformed into planetesimals. The dotted grey line displays the midplane dust-to-gas ratio at the dust peak $r_{\rm peak}$, which instead follows a linear scale from 0 to 1.}
    \label{fig:mass_stat_grid}
\end{figure*}

\begin{figure*}
    \centering
    \includegraphics[width=\linewidth]{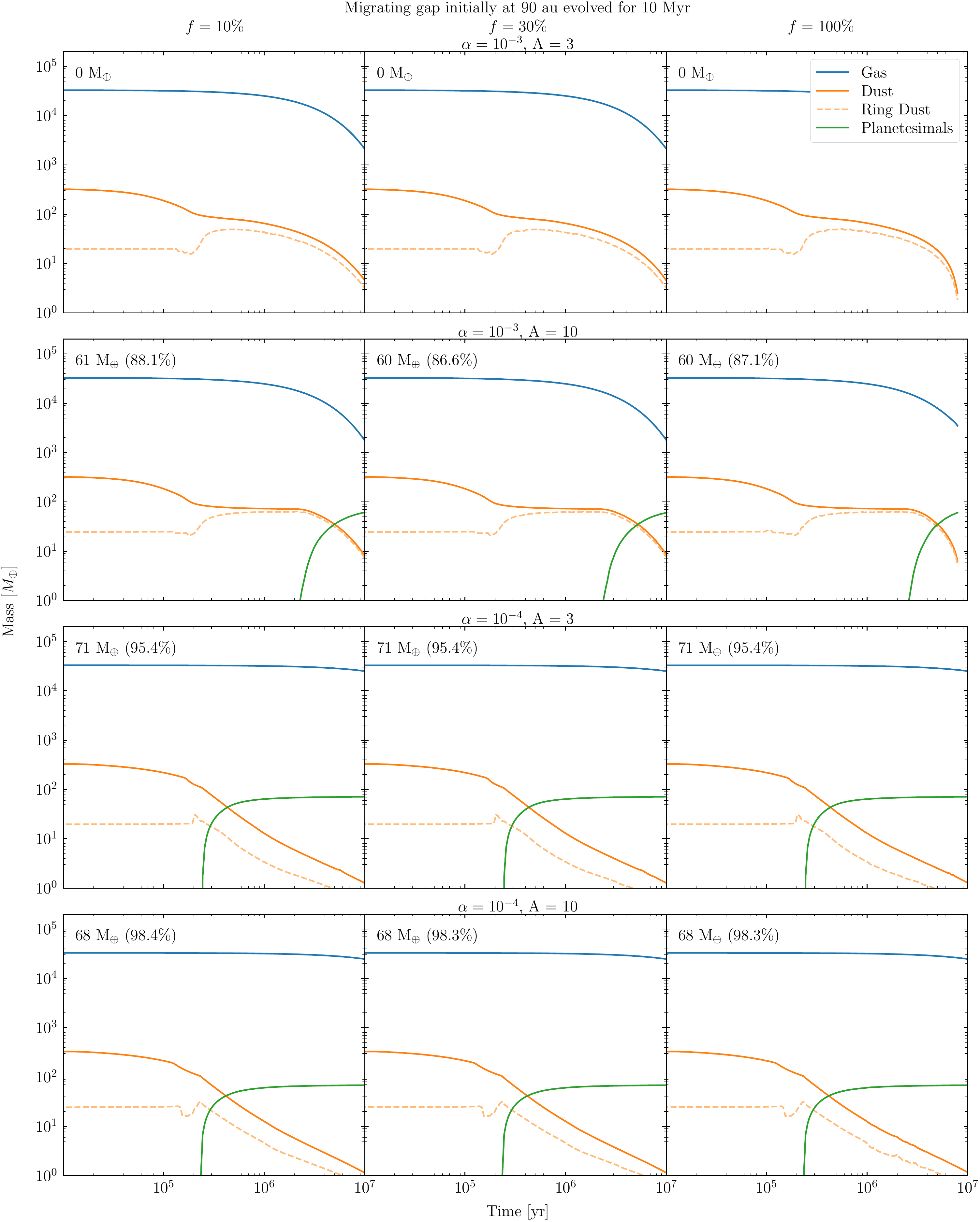}
    \caption{Mass evolution of the gas, dust, ring dust and planetesimals in a protoplanetary disc with a gap initially at 90 au migrating at $f$ = 10, 30 and 100\% of the nominal velocity with varying viscosity parameter$\alpha$ and gap amplitude $A$. If the gap is at $x = -d$ and dust peak is at $x = 0$, the ring dust was computed by integrating over $x = \pm d/2$. The final planetesimal mass for each simulation is presented, adjacent to the final percentage of initial exterior dust mass transformed into planetesimals.}
    \label{fig:mass_mov_grid}
\end{figure*}

The mass evolution of the system for the stationary and migrating gap simulations are shown in Figure~\ref{fig:mass_stat_grid} and Figure~\ref{fig:mass_mov_grid} respectively. See \S\ref{sec:results} for the corresponding discussion.

\section{Dust Mass Distribution}
\label{appendix_dist}
\begin{figure*}
    \centering
    \includegraphics[width=\linewidth]{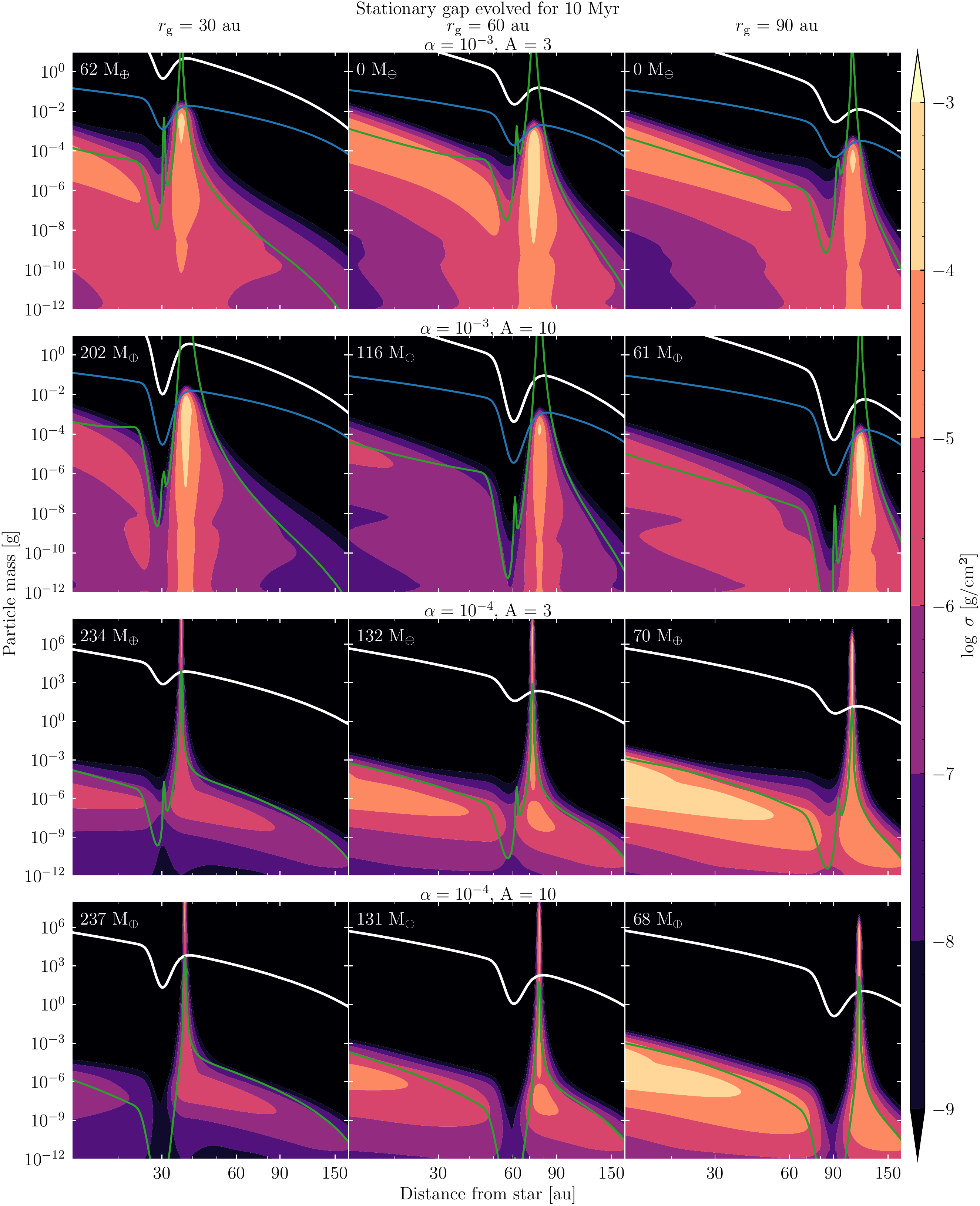}
    \caption{Dust mass distribution of a protoplanetary disc with a stationary gap at $r_{\rm g}$ = 30, 60 and 90 au with varying viscosity parameter $\alpha$ and amplitude $A$ evolved for 10 Myr. The colour bar $\sigma$ represents the dust surface density. The white line corresponds to dust particles with a Stokes number of 1. The blue and green lines are the dust fragmentation and drift limits respectively.}
    \label{fig:dist_stat_grid}
\end{figure*}

\begin{figure*}
    \centering
    \includegraphics[width=\linewidth]{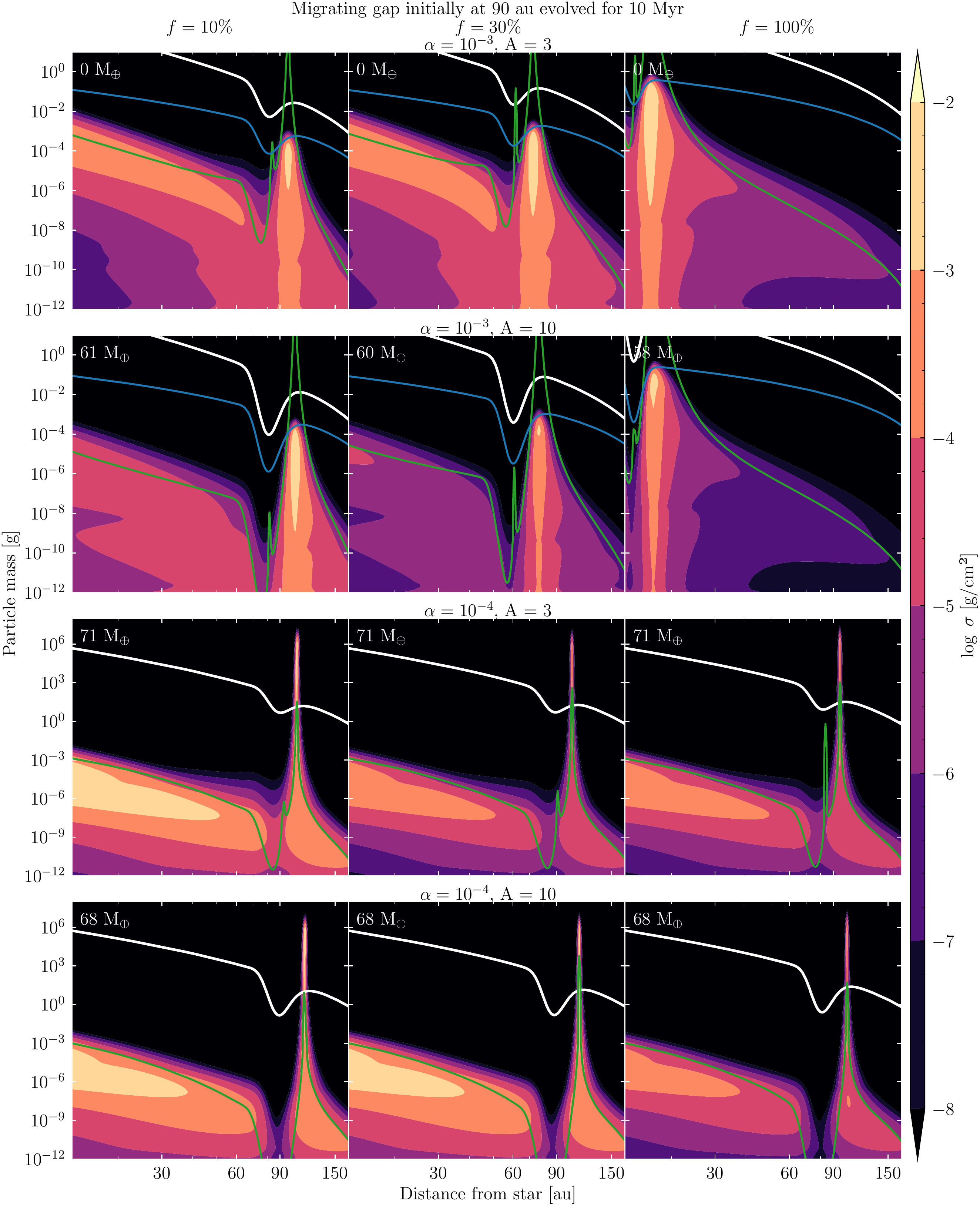}
    \caption{Dust mass distribution of a protoplanetary disc with a gap initially at 90 au migrating at $f$ = 10, 30 and 100\% of the nominal velocity with varying viscosity parameter $\alpha$ and amplitude $A$ evolved for 10 Myr. The colour bar $\sigma$ represents the dust surface density. The white line corresponds to particles with a Stokes number of 1. The blue and green lines are the dust fragmentation and drift limits respectively.}
    \label{fig:dist_mov_grid}
\end{figure*}

The dust (particle) mass distribution of the system at the final epoch for the stationary and migrating gap simulations are shown in Figure~\ref{fig:dist_stat_grid} and Figure~\ref{fig:dist_mov_grid} respectively. 

\bsp	
\label{lastpage}
\end{document}